\documentclass[twoside]{article}

\usepackage[accepted]{aistats2023}
\usepackage{natbib}
\usepackage{hyperref}       
\usepackage{url}            

\newcommand{\lp}{\left(}
\newcommand{\rp}{\right)}
\newcommand{\lb}{\left[}
\newcommand{\rb}{\right]}
\newcommand{\lbp}{\left\{}
\newcommand{\rbp}{\right\}}
\newcommand{\lba}{\left\lvert}
\newcommand{\rba}{\right\rvert}
\newcommand{\lV}{\left\lVert}
\newcommand{\rV}{\right\rVert}
\newcommand{\mv}{\middle\vert}

\newcommand{\mcal}{\mathcal}

\newcommand{\bbm}{\mathbbm}
\newcommand{\mbb}{\mathbb}
\newcommand{\msf}{\mathsf}

\newcommand{\ra}{\rightarrow}

\newcommand{\lan}{\langle}
\newcommand{\ran}{\rangle}

\newcommand{\eqDef}{\triangleq}
\newcommand{\diid}{\overset{\text{i.i.d.}}{\sim}}

\newcommand{\E}{\mathbb{E}}
\newcommand{\Var}{\mathsf{Var}}

\renewcommand{\Pr}{\mathbb{P}}

\newcommand{\eps}{\varepsilon}

\usepackage{float}
\usepackage{wrapfig}
\usepackage{bm, bbm}
\usepackage{mathtools}
\usepackage{amsmath, amssymb, mathrsfs}
\usepackage[ruled,vlined]{algorithm2e}
\usepackage[shortlabels]{enumitem}
\usepackage{color}

\newtheorem{lemma}{Lemma}[section]
\newtheorem{theorem}{Theorem}[section]

\newtheorem{corollary}{Corollary}[section]
\newtheorem{definition}{Definition}[section]
\newtheorem{remark}{Remark}[section]
\newtheorem{claim}{Claim}[section]

\begin{document}

%

%

\twocolumn[

\aistatstitle{The communication cost of security and privacy in federated frequency estimation}

\aistatsauthor{ Wei-Ning Chen \And Ayfer \"Ozg\"ur \And  Graham Cormode \And Akash Bharadwaj}

\aistatsaddress{ Stanford University \And  Stanford University \And Meta AI \And Meta AI} ]

\begin{abstract}
 We consider the federated frequency estimation problem, where each user holds a private item $X_i$ from a size-$d$ domain and a server aims to estimate the empirical frequency (i.e., histogram) of $n$ items with $n \ll d$. Without any security and privacy considerations, each user can communicate its item to the server by using $\log d$ bits. A naive application of secure aggregation protocols would, however, require $d\log n$ bits per user. Can we reduce the communication needed for secure aggregation, and does security come with a fundamental cost in communication?  
 
 In this paper, we develop an information-theoretic model for secure aggregation that allows us to characterize the fundamental cost of security and privacy in terms of communication. We show that with security (and without privacy) $\Omega\left( n \log d \right)$ bits per user are necessary and sufficient to allow the server to compute the frequency distribution. This is significantly smaller than the $d\log n$ bits per user needed by the naive scheme, but significantly higher than the $\log d$ bits per user needed without security. To achieve differential privacy, we construct a linear scheme based on a noisy sketch which locally perturbs the data and does not require a trusted server (a.k.a. distributed differential privacy). We analyze this scheme under $\ell_2$ and $\ell_\infty$ loss. By using our information-theoretic framework, we show that the scheme achieves the optimal accuracy-privacy trade-off with optimal communication cost, while matching the performance in the centralized case where data is stored in the central server.
 
\end{abstract}
\allowdisplaybreaks
\section{Introduction}
Modern data is increasingly born at the edge and can carry sensitive user information. To make efficient use of this data while protecting individual information from being revealed to the public or service providers, in recent years there has been a strong desire for data science methods that allow servers to collect population-level information from a set of users without knowing each individual value. 
Consider, for instance, frequency estimation which serves as a fundamental building block for many analytics tasks. Each user holds an item $X_i$ from a size-$d$ domain $\mcal{X}$, and the server aims to learn the empirical frequency (i.e., the histogram) of all items. Can the server learn the empirical frequency distribution of the items without learning each individual's item?

Recently, distributed protocols based on multi-party computation (MPC) such as secure aggregation (SecAgg)~\citep{bonawitz2016practical} have emerged as a powerful tool to securely aggregate population-level information from a set of users. 
In particular, SecAgg allows a single server to compute the population sum (and hence also the average) of local variables (often vectors), while also ensuring no additional information, other than the sum, is released to the server or other participating entities. This can be achieved, for example, by having users apply additive masks on their local vectors which cancel out upon addition at the server. SecAgg is widely used within protocols for secure federated learning and secure statistics gathering, which both rely on vector summation.

A straightforward way to use SecAgg for the empirical frequency estimation problem above is to have each user represent their item $X_i$ as a $d$-dimensional one-hot vector (i.e., a vector with a single $1$ in the $X_i$-th coordinate and zero otherwise), so that the sum of all one-hot vectors (which is revealed to the server by SecAgg) gives the desired histogram. However, this requires  $d\log n$ bits of communication per user since each user has to communicate a masked vector of dimension $d$ (with each entry taking values in a finite field of size $n$). This is a drastic increase from the $\log d$ bits per user needed to communicate each item in the absence of any security considerations. 

Can we reduce the communication cost of secure aggregation, and does security come with a fundamental cost in communication? This is the main question we investigate in this paper. We show:
\begin{itemize}[leftmargin=1.5em, topsep=0em]\itemsep0em
    \item The communication cost for secure frequency estimation can be reduced from $O(d\log n)$ to $O(n\log d)$ when $n \ll d$.  This is the relevant regime in many real-world applications (e.g., location tracking~\citep{bagdasaryan21sparse}, language modeling, web-browsing, etc.) where $d$ can be very large and computational constraints limit the number of users that can participate in each SecAgg round.
    \item Complementarily, any aggregation protocol that is information-theoretically secure needs $\Omega(n\log d)$ bits per user to perfectly recover the histogram. 
    To show this we develop an information-theoretic model for secure aggregation and prove a lower bound on the communication cost of any secure aggregation protocol.
\end{itemize}
This reveals that while the communication cost of secure frequency estimation can be reduced with more carefully designed schemes (e.g., we show that one can formulate it as an $\ell_1$ constrained integer linear inverse problem), there is a fundamental price to computing the histogram \emph{securely}: in the absence of any security considerations, each user needs $\log d $ bits, and hence the total communication cost for \textit{all} users is $n\log d$; with security, each user \textit{individually} incurs the $n\log d$ bits communication cost (i.e., $O(n^2 \log d)$ in total).

Secure aggregation alone does not provide any provable privacy guarantees such as differential privacy (DP). Sensitive information may still be revealed from the aggregated population statistics, causing potential privacy leakage. To address this issue, a common approach is to  perturb the aggregated information by adding noise before passing it to downstream analytic tasks. With a privacy requirement,  the empirical frequency can be estimated only approximately, with an amount of distortion that depends on the privacy level, number of participating users, and the loss function. This distortion due to DP also allows for some `slack' in the secure aggregation framework -- as long as secure aggregation returns an approximate sum with a distortion small enough compared to the distortion due to DP, we can achieve order-wise the same performance as with only the DP constraint. This observation leads us to study the communication cost of secure aggregation for computing an \emph{approximate} sum rather than an exact sum of the user values. We show that computing an approximate sum requires less communication, and the optimal communication cost can be characterized by a rate-distortion function that depends on the error and the loss function. While security drastically increases the communication cost, we show that privacy helps us reduce it.

Our end goal is to arrive at secure and private frequency estimation protocols that provide differential privacy guarantees without putting trust in the service provider, while at the same time achieving the optimal privacy-accuracy-communication trade-off. To this end, we develop a \emph{user-level} DP protocol for frequency estimation, where users compute a summary of their local data, perturb these slightly, and employ SecAgg to simulate some of the benefits of a trusted central party. The untrusted server has access only to the aggregated reports with the aggregated perturbations.  We show that the end-to-end privacy-accuracy trade-off achieved by this scheme is optimal and matches the trade-off achievable with a trusted server, i.e., in a centralized setting where the server receives the data as it is and perturbs it after aggregation. Furthermore, by using our aforementioned information-theoretic framework for securely computing an approximate sum, we show that the communication cost of this scheme is also optimal.

\textbf{Our contributions.}
The main contributions of our paper can be summarized as follows:
\begin{itemize}[leftmargin=1.5em, topsep=0em]

    \item We provide an information-theoretic view on secure aggregation and analyze the amount of communication needed for securely computing the sum either exactly or approximately. In the case of exact recovery, we show that the per-user communication cost is lower bounded by the entropy of the sum; for  approximate recovery under a general loss function $\ell(\cdot)$, we specify the communication-distortion trade-offs.
    
    \item We specialize these information-theoretic lower bounds to frequency estimation with and without differential privacy constraints. We show that without privacy $\Omega(n\log d)$ bits per user are needed to allow the server to learn the exact histogram. We also characterize the minimal communication cost when differential privacy is required.
    
    \item We introduce schemes that match the above information-theoretic communication lower bounds. In particular, we show that to perfectly recover the exact histogram (without privacy), one can achieve the optimal $O(n\log d)$ bits per-user communication by applying SecAgg and solving a linear inverse problem. To achieve differential privacy, we construct a linear scheme based on noisy sketch (with proper modifications tailored to the specific loss function) which locally perturbs the data and does not require a trusted server (a.k.a user-level DP). We show that this scheme achieves the (nearly) optimal accuracy-privacy trade-off with optimal communication cost, while matching the performance in the centralized case where data is stored in the central server.
\end{itemize}

\textbf{Organization.}
The rest of the paper is organized as follows. We discuss the related works in Section~\ref{sec:related_works}. In Section~\ref{sec:secagg}, we introduce a general framework for SecAgg and the corresponding information-theoretic security it provides and proves general communication lower bounds on computing the exact or approximate sum. In Section~\ref{sec:secure_fe}, we apply SecAgg to the frequency estimation problem and specify the optimal communication cost. Finally, in Section~\ref{sec:secure_private_fe}, we incorporate the differential privacy constraint and characterize the optimal privacy-communication-accuracy trade-offs.

\textbf{Notation.}
Throughout this paper, we use $[m]$ to denote the set of $\lbp 1,...,m \rbp$ for any $m \in \mbb{N}$. Random variables (vectors) $(X_1,...,X_m)$ are denoted as $X_{[m]}$ or $X^m$. We also make use of Bachmann-Landau asymptotic notation, i.e., $O, o, \Omega, \omega, \text{ and } \Theta$. We use $H(X)$ (or $H(P_X)$) to denote the Shannon entropy of $X$ with base 2. Finally, for random variables $X, Y$, $I(X; Y)$ denotes the mutual information, i.e., $I(X; Y) \eqDef \E_{X}\lb D_{\msf{KL}}\lp P_{Y|X}\Vert P_Y\rp \rb$.

\section{Related Work}\label{sec:related_works}
\textbf{Secure aggregation.}  Single-server SecAgg is a cryptographic secure multi-party computation (MPC) that enables users to submit vector inputs, such that the server learns just the sum of the users' vectors. This is usually achieved via additive masking over a finite group \citep{bonawitz2016practical, bell2020secure}. The single-server setup makes SecAgg particularly suitable for federated learning \citep{kairouz2021distributed, agarwal2021skellam} or federated analytics \citep{choi2020differentially}, and a recent line of works \citep{jahani2022swiftagg+, so2021turbo, choi2020communication, kadhe2020fastsecagg, yang2021lightsecagg} aim to scale it up by improving the communication or computation overhead. However, all of the above works focus on a general setting where the local vectors can be arbitrary; meanwhile, in the frequency estimation problem with large domain size, local vectors are one-hot and the histogram is typically sparse. Without secure aggregation such sparsity can be leveraged to reduce the communication cost \cite{acharya2019hadamard, han2018distributed,barnes2019lower, acharya2019inference, acharya2019inference2, acharya2020estimating, chen2021breaking, chen2021pointwise}. However, with secure aggregation, it is not clear if and how sparsity can be leveraged to reduce communication, which is the main focus of our work. 


\textbf{Differential Privacy.} To achieve provable privacy guarantees SecAgg is insufficient as even the sum of local model updates may still leak sensitive information \citep{melis2019exploiting, song2019auditing, carlini2019secret, shokri2017membership} and so differential privacy (DP) \citep{dwork2006our} can be adopted. 
By having the noise added locally and letting the server aggregate local information via SecAgg, the DP guarantees do not rely on users' trust in the server. 
This \emph{user-level DP} (also referred to as distributed DP in the literature) framework has recently been adopted in private federated learning\citep{agarwal2018cpsgd, kairouz2021distributed, agarwal2021skellam, chen2022fundamental}. 
In this work, we use the Poisson-binomial mechanism as a primitive \citep{chen2022poisson} to achieve user-level DP.

We also distinguish our setup from the local DP setting \citep{kasiviswanathan2011can, evfimievski2004privacy, warner1965randomized}, where the data is perturbed on the user-side before it is collected by the server. 
Local DP, which allows for a possibly malicious server, is stronger than distributed DP, which assumes an honest-but-curious server. Consequently, local DP suffers from worse privacy-utility trade-offs~\citep{duchi2013local, ye2017optimal, barnes2020fisher, acharya2021inference}.

SecAgg can be viewed as a privacy amplification technique that amplifies weak local DP to much stronger central DP guarantees. 
Other amplification techniques are based on different cryptographic techniques such as secure shuffling \citep{erlingsson2019amplification, balle2019privacy, balle2020private, balcer2019separating} or distributed point functions \citep{gilboa14dpf}. 
While the fundamental communication cost for SecAgg that we show in our paper can be potentially circumvented by these methods, these techniques either require the existence of a trusted shuffler or assume multiple servers that do not collude.

\textbf{Private frequency estimation and heavy hitters.}
Private frequency estimation, a.k.a. histogram estimation, is a canonical task that has been heavily studied in the DP literature \citep{dwork2006calibrating}. 
When subject to $\ell_\infty$ loss, it is the same as the heavy hitter problem. Under the centralized setting, typical techniques for releasing a private histogram include the addition of noise (and thresholding the counts) \citep{dwork2006calibrating, ghosh2012universally, korolova2009releasing, bun2016concentrated, balcer2017differential} or sampling-and-thresholding \citep{zhu2020federated, cormode2022sample}. 
The private heavy hitter problem has also been heavily studied under the local or multiparty DP model \citep{Bassily2015, Bassily2017, bun18hh, bun2019heavy, huang2021frequency}. Our work, however, is under the user-level DP model, under which most previous techniques cannot be directly applied. Our privatization technique makes use of noisy count-sketch, which is close to the work of \cite{choi2020differentially}, in which a general distributed noisy sketch framework is analyzed. In this work, we use a similar technique to characterize the exact communication cost and show that a noisy sketch can achieve the communication lower bound.

\section{Secure Aggregation}\label{sec:secagg}
In this section, we formulate a general framework for secure aggregation with a single server and $n$ users. 
Assume each user $i\in [n]$ holds local information (as a vector) $X_i \in \mcal{X}$, and the server aims to compute the sum $\mu(X_1, X_2,...,X_n) \eqDef \sum_{i \in [n]} X_i$. During the aggregation, up to $D$ clients may drop out, and the secure aggregation protocol should still be able to recover the sum of the remaining clients. In general, an aggregation protocol consists of encoding functions $g_i,\, i\in[n]$ at the users and an aggregation function $f$ at the server such that:

\begin{enumerate}[leftmargin = 1.5em, topsep=0pt]
\setlength\itemsep{0em}
    \item Each user encodes their local information $X_i$  as $Y_i = g_i(X_i; \theta_i)$, where $\theta_i$ is randomness available at the $i$'th user which is independent of $X_i$ but may depend on other $\theta_{i'}$ for $i' \in [n]\setminus i$. 
    \item The server observes $Y_i$ for $i\in[n]\setminus\mathcal{D}$, i.e., the messages of the available users. If $\mathcal{D}=\emptyset$, i.e. there are no dropouts, it estimates the sum $\mu(X^n)$ by a (deterministic) function $f\lp Y^n\rp$.
    
    \item  If $\mathcal{D}\neq\emptyset$, the server invokes a second round of communication with the surviving clients to recover the masks of the dropout users. In this round, the server collects $h\left( \theta_{[n]}, \mathcal{D} \right)$, where $h(\cdot)$ is a general function of local secrets $\theta_{[n]}$, and uses the information it collects  over the two rounds, $Y_{[n]\setminus\mathcal{D}}$ and $h\left( \theta_{[n]}, \mathcal{D} \right)$, to estimate the sum of the surving clients $\mu\left( X_{[n]\setminus \mathcal{D}} \right)$.
\end{enumerate}

\begin{figure}
    \centering
    \includegraphics[width=0.6\linewidth]{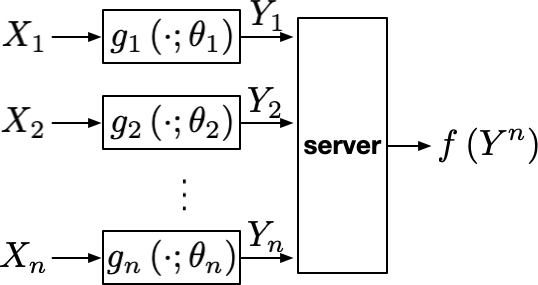}
    \caption{A framework for SecAgg (illustrated for the case without dropouts).}
    \label{fig:secagg_itform}
\end{figure}

\textbf{Security constraints:}
We call the aggregation protocol that can tolerate $D$ drop-outs \emph{secure}, if it satisfies the following two conditions on mutual information for any distribution $P_{X^n}$ imposed on the user data:
\begin{align}
    &\forall \mathcal{D} \subseteq [n], 
     I\left( Y_{[n]\setminus \mathcal{D}}, h\left( \theta_{[n]}, \mathcal{D} \right);  X_{[n]\setminus \mathcal{D}} \middle\vert \mu\left( X_{[n]\setminus \mathcal{D}} \right)\right) = 0 \tag{S1}\label{S1}, \\
     & \forall |\mathcal{D}| > D, \, I\left( Y_{[n]\setminus \mathcal{D}}, h\left( \theta_{[n]}, \mathcal{D} \right); X_{[n]\setminus \mathcal{D}} \right) = 0. \tag{S2}\label{S2}
\end{align}

\eqref{S1} implies that $X_{[n]\setminus \mathcal{D}}-\mu\lp X_{[n]\setminus \mcal{D}} \rp - (Y_{[n]}, h)$ forms a Markov chain,
and hence given $\mu\left( X_{[n]\setminus \mathcal{D}} \right)$ the server  cannot deduce any further information about $X_{[n]\setminus \mcal{D}}$ from the information it gathers over the two stages of the scheme, $Y_{[n]\setminus \mathcal{D}}$ and $h\left( \theta_{[n]}, \mathcal{D} \right)$;
\eqref{S2} states that without a sufficient number of users participating (e.g., when $|\mcal{D}| \geq D$), the server cannot learn any information about the user data. 

 Note that the same framework can be used to include colluding users by allowing $h(\cdot)$ to contain information about both the masks $\theta_D$ and the information $X_{\mcal{D}}$ of the users in $\mcal{D}$, i.e. $h\left( \theta_{[n]}, X_{\mcal{D}},\mathcal{D} \right)$. Security for the remaining users is ensured with the same constraints \eqref{S1} and \eqref{S2}.

The two security requirements above are satisfied by most practical secure aggregation protocols such as\citet{bonawitz2016practical} and \citet{bell2020secure}. In the next section, we show that these security requirements come at a fundamental and significant communication cost.

\textbf{Correctness constraints:}

In the absence of any privacy considerations, we impose the following correctness requirement on the protocol, which ensures that it always outputs the correct sum:
\begin{equation}
\forall |\mathcal{D}| \leq D,\,\Pr\left\{ f\left( Y_{[n]\setminus \mathcal{D}}, h\left( \theta_{[n]}, \mathcal{D} \right) \right) =  \mu\left( X_{[n]\setminus \mathcal{D}} \right) \right\} = 1.
\tag{C1}\label{C1}
\end{equation}
We are also interested in the case where the server recovers the sum approximately under a certain loss function (this is the relevant setting under differential privacy constraints). 
Let $\ell(\cdot, \cdot)$ be a loss function defined on the domain of $\mu\lp X^n \rp = \sum_{i=1}^n X_i$. We refer to the following approximate recovery criterion as the $\beta$-distortion criterion:
\begin{equation}\tag{C1$^\prime$} \label{C1prime}
    \forall |\mathcal{D}| \leq D,\,\mathbb{E}\left[ \ell \left( f\left( Y_{[n]\setminus \mathcal{D}}, h\left( \theta_{[n]}, \mathcal{D} \right) \right),  \mu\left( X_{[n]\setminus \mathcal{D}} \right)\right) \right] \leq \beta.
\end{equation}
Note that under this criterion the server recovers the sum with distortion $\beta$ under the loss function $\ell$.

\textbf{Communication cost:} The communication cost of an aggregation protocol for user $i$ is given by $\max_{P_{X^n}} H(Y_i)$ (i.e., the worst-case entropy for any possible joint distributions over the local data). This is the (maximum over the choice of $X^n$) number of bits node $i$ needs on average to communicate $Y_i$ using an optimal compression scheme. 

\subsection{Communication Lower Bounds}\label{subsec:general_lower_bounds}
Next, we present general communication lower bounds on estimating the sum of $n$ random variables $X_{[n]}$ (where we do not make any assumptions on the domain of $X_i$) under the security constraints \eqref{S1} and \eqref{S2}.  

\begin{lemma}[Lower bound for perfect recovery]\label{lemma:noiseless_lower_bound}
Let $\mathcal{D} \subset [n]$ be the set of dropout clients, such that $|\mathcal{D}| \leq D$ for some $D \leq \frac{n}{2}$. Under the correctness constraint \eqref{C1} and security constraints \eqref{S1} and \eqref{S2} on the protocol, it holds that for all $i \in [n]\setminus\mathcal{D}$, $H(Y_i) \geq H\left( \sum_{i\in[n]\setminus \mathcal{D}} X_i \right),$
where $H(\cdot)$ is the Shannon entropy.
\end{lemma}

Note that $H\left( \sum_{i\in[n]\setminus \mathcal{D}} X_i \right)$ quantifies the information the server is able to learn about the user data. The lemma states that in a secure protocol, the entropy of each individual message should be at least as large the total information communicated to the server.  In the following lemma, we characterize how this lower bound is modified when the  server needs to recover the sum only approximately.  
\begin{lemma}\label{lemma:lossy_lower_bound}

Let $n' \eqDef n - |\mathcal{D}|$. Let $\ell(\cdot, \cdot)$ be a loss function defined on the domain of $ \mu\left( X^{n'} \right) = \sum_{i\in[n]\setminus \mathcal{D}} X_i$. Under the $\beta$-approximate recovery criterion \eqref{C1prime} and the security constraints \eqref{S1} and \eqref{S2}, it holds that 
$$ H(Y_i) \geq R(\beta) \text{, for all } i \in [n],$$
where $R(\beta)$ is the solution of the following rate-distortion problem:
{\small
\begin{align}\label{eq:rate_function}
&R\left( \beta \right) \eqDef 
    &\begin{pmatrix}
     \min & I\left( Y^{n'} ;  \mu\left( X^{n'} \right)\right)\\
     \textrm{s.t. } &\min_{\hat{\mu}} \mathbb{E}\left[ \ell\left( \hat{\mu}\left( Y^{n'}  \right),  \mu\left( X^{n'} \right) \right) \right] \leq \beta
    \end{pmatrix}
\end{align}}
where the first minimization is taken over all conditional probability $P_{Y^{n'} |\mu\left( X^{n'} \right)}$.

\end{lemma}

Lemma~\ref{lemma:lossy_lower_bound} suggests that under the $\beta$-approximate recovery criterion, the communication load of a secure aggregation protocol is  lower-bounded by $R(\beta)$ per user. In Section~\ref{sec:secure_private_fe}, we explicitly characterize $R(\beta)$ for the frequency estimation problem. 

\section{Secure Frequency Estimation}\label{sec:secure_fe}
In this section, we formally define the frequency estimation problem with security constraints \eqref{S1} and \eqref{S2} and study the optimal communication cost. Assume each user $i$ holds an item $X_i$ in a size $d$ domain $\mcal{X}$ and the server aims to estimate the histogram of the $n$ items. Let $X_i \in \mcal{X} \eqDef \lbp e_1,...,e_d \rbp \in \lbp 0, 1\rbp^d$, i.e., each item is expressed as a one-hot vector. Note that this is without loss of generality since the encoding functions $g_i$ at the users can be arbitrary. 
Then, the histogram of the $n$ items can be expressed as $\mu\lp X^n \rp \eqDef \sum_{i \in [n]} X_i \in [n]^d$. We mainly focus on the high-dimensional regime where $d \gg n$, and our goal is to characterize the communication needed to securely compute $\mu(X^n)$. 

To this end, we first apply the (general) lower bound derived in Section~\ref{subsec:general_lower_bounds} with $X_i \in \lbp e_1, e_2, ...,e_d \rbp$. For simplicity, we present our results without dropouts (i.e., $\mcal{D} = \emptyset$), but extending to the $|\mcal{D}|>0$ case is immediate. Our lower bound is obtained by imposing a worst-case prior distribution on $X^n$ we arrive at the following corollary:
\begin{corollary}\label{cor:exact_lower_bound}
Let $X_i \in \lbp e_1,...,e_d \rbp$ for $i \in [n]$. Under the same set of constraints as in Lemma~\ref{lemma:noiseless_lower_bound}, there exists a worst-case prior distribution $\pi_{X^n}$ such that  
\begin{equation}\label{eq:worst_case_noiseless_lower_bound}
\textstyle
    H(Y_i) \geq H\lp \sum_{i=1}^n X_i\rp =\Omega\lp n \log d \rp,
\end{equation}
where the entropy $H\lp \sum_{i=1}^n X_i\rp$ is computed with respect to $X^n \sim \pi_{X^n}$.
\end{corollary}

In the rest of this section, we outline a communication-efficient secure frequency estimation scheme based on solving a linear inverse problem, and the resulting per-user communication cost matches the lower bound in the corollary. 
We state this result, together with the lower bound in Corollary~\ref{cor:exact_lower_bound}, as our main theorem:
\begin{theorem}
To securely (i.e., under \eqref{S1} and \eqref{S2}) and correctly (i.e., under \eqref{C1}) compute the histogram from $n$ users, it is both sufficient and necessary for each user to send $\Theta\lp n \log d \rp$ bits to the server.
\end{theorem}

\subsection{Reducing Communication via Sparse Recovery}\label{subsec:perfect_recovery_achievability}

In this section, we propose a scheme that shows that the communication cost can be reduced to the information theoretic $\Omega\lp n \log d \rp$ bits lower bound. 
Our scheme depends on two main ingredients: (1) a specific construction of a secure aggregation protocol, often called SecAgg, due to \cite{bonawitz2016practical}, and (2) a linear binary compression scheme based on random coding. For simplicity, we describe our schemes for the case of no dropouts, but our schemes can be readily extended to handle dropouts or colluding users since they are based on SecAgg (which is designed to tolerate dropouts/colluding users). 

In a nutshell, the encoding steps of SecAgg \citep{bonawitz2016practical} consist of (i) mapping $X_i$ into an element of a finite group (where, without loss of generality, we assume the group is $\mbb{Z}^m_M$ for some $m, M \in \mbb{N}$), and then (ii) adding a random mask $\theta_i \in \mbb{Z}^m_M$ so that $Y_i = \mcal{A}_{\msf{enc}}(X_i) + \theta_i$. The mask $\theta_i$ has uniform marginal density, is independent of $X^n$, and satisfies $\sum_{i\in[n]} \theta_i = 0$. 
Upon receipt of $Y^n$, the server computes the sum of $Y^n$ and decodes it via $\mcal{A}_{\msf{dec}}\lp \sum_{i} \mcal{A}_{\msf{enc}}(X_i) \rp$. The goal is to design mappings $\lp \mcal{A}_{\msf{enc}}, \mcal{A}_{\msf{dec}}\rp$, so that 
\begin{itemize}[topsep=0em, leftmargin=1.5em]
    \item the outcome correctly recovers $\mu\lp X^n \rp$, i.e., $\mcal{A}_{\msf{dec}}\lp \sum_{i} \mcal{A}_{\msf{enc}}(X_i) \rp = \sum_{i=1}^n X_i$;
    \item the per-user communication cost $m \log M$ is minimized.
\end{itemize}

Due to the linearity of SecAgg (i.e., the server obtains the sum of $\mcal{A}_{\msf{enc}}$), $\mcal{A}_{\msf{enc}}$ is usually constructed via a linear mapping, so that $\mcal{A}_{\msf{enc}}(X_i) \eqDef S\cdot X_i$ for some $S \in \lp\mbb{Z}_M\rp^{m\times d}$. In this case, the sum of the encodings is the same as the encoding of the sum, i.e.,
\begin{equation}\label{eq:secagg} 
\textstyle
\sum_i \mcal{A}_{\msf{enc}}\lp X_i \rp = \mcal{A}_{\msf{enc}}\lp \sum_i X_i \rp = S \mu\lp X^n \rp.
\end{equation}
To recover $\mu$ from $S\mu$, the server solves a linear inverse problem, which has a unique solution only if $S$ is ``invertible'' for all possible $\mu$'s. For example, a naive choice of $S$ can be the identity mapping $I_d$, which encodes each $X_i$ as a one-hot vector. In this case, the size of the finite group $\mbb{Z}^m_M$ is $(M, m) = (n, d)$, and the communication complexity is $d \log n$ bits. This is far from the lower bound $\Omega\lp n \log d\rp$ when $n \ll d$.

Can we design a better embedding matrix $S$ with smaller range (i.e., with smaller $(M, m)$) than the naive choice $I_d$ so that $y = S\mu$ is solvable? Specifically, define $\mcal{H}_n$ to be the collection of all possible $n$-histogram, i.e., 
$\mcal{H}_n \eqDef \lbp \mu \in \mbb{Z}_+^d \mv \lV \mu \rV_1 = n \rbp. $
Our goal is to show that there exists an $S \in \{0, 1\}^{m\times d}$ with $m= O\lp n\log d/\log n\rp$,  such that $y = S\mu$ is solvable for all $\mu \in \mcal{H}_n$. Using such $S$ as our local embedding, the resulting communication cost becomes $O(n\log d)$ and hence matches the lower bound. We summarize this in the following theorem
\begin{theorem}\label{thm:l1_linear_inverse}
Let $\mcal{H}_n$ be the collection of all valid $n$-histograms formally defined as above. Then there exists an embedding matrix  $S \in \{0, 1\}^{m\times d}$ with $m= O\lp \frac{n\log d}{\log n}\rp$, such that
\begin{equation}\label{eq:S_invertible}
    \forall  \mu_1, \mu_2 \in \mcal{H}_n, \, \mu_1 \neq \mu_2 \Longrightarrow S\mu_1 \neq S\mu_2.
\end{equation}
\end{theorem}

Theorem~\ref{thm:l1_linear_inverse} can be viewed as a generalization of classical (non-adaptive) Quantative Group Testing (QGT) \citep{bshouty2009optimal, wang2016extracting, scarlett2017phase, gebhard2019quantitative}, in which the linear inverse problem is defined over the $\ell_1$ constrained \emph{binary} vectors $\mcal{G}_n \eqDef \lbp \nu \in \{0, 1\}^d\, \mv\, \lV \nu \rV_0 = n \rbp$. To prove the existence of such $S$, we follow the idea of \citet{wang2016extracting} by constructing $S$ in a probabilistic way, i.e., generating each element of $S$ as an independent $\msf{Bern}(1/2)$ random variable. We then show that as long as $m = \Omega\lp n \log d / \log n \rp$, \eqref{eq:S_invertible} holds with high probability, hence concluding the existence of $S$. One key step that generalizes the result from classical QGT is an application of Sperner's theorem \citep{sperner1928satz, lubell1966short}, which may be of independent interest. The proof of Theorem~\ref{thm:private_secure_fe_achievability} can be found in Appendix~\ref{proof:l1_linear_inverse}.

\paragraph{Comparison to compressed sensing.} Note that as the set of $n$-histograms is a subset of $n$-sparse vectors in $\mbb{R}^d$, it may be tempting to use standard sparse recovery techniques such as compressed sensing \citep{donoho2006compressed, donoho2006most} (e.g., with the classical Rademacher ensemble, see \cite[Chapter~7]{wainwright2019high}). This can allow us to reduce the dimensionality from $d$ to $m = O(n\log d)$. However, each coordinate of the embedded vector can range from $-n$ to $n$ (using the Rademacher ensemble), and requires $O(\log n)$ bits to represent it and  the total communication cost is $O(n\log d\cdot \log n)$ leading to an extra $\log n$ factor. Theorem~\ref{thm:l1_linear_inverse} is necessary in order to obtain a information-theoretically optimal solution.

On the other hand, the scheme proposed in the proof of Theorem~\ref{thm:l1_linear_inverse}, though optimal in terms of communication efficiency, is computationally infeasible. 
It requires exhaustively scanning over  $\mcal{H}_n$ to find the unique consistent histogram $\mu^*$, and hence the computation cost is $\Omega\lp d^n \rp$. It remains open if one can design computationally efficient schemes (e.g., a scheme with computational cost $\msf{poly}(n, d)$ or ideally $\msf{poly}(n, \log d)$) that achieves the best $O(n\log d)$ communication cost.

\section{Secure and Private Frequency Estimation}\label{sec:secure_private_fe}
Secure aggregation alone does not provide any privacy guarantees. In this section, we study the private frequency estimation problem, in which, apart from security constraints \eqref{S1} and \eqref{S2}, we also impose a privacy constraint on our protocol. 
Our goal is to characterize the communication required for the optimal accuracy-privacy tradeoff.
We first state the definition of differential privacy~\citep{dwork2006calibrating}.

\begin{definition}[Differential Privacy (DP)]\label{def:DP}
For $\varepsilon, \delta \geq 0$, a randomized mechanism $M$ satisfies $(\varepsilon, \delta)$-DP if for all neighboring datasets $D, D'$ and all $\mcal{S}$ in the range of $M$, we have that 
$$ \Pr\lp M(D) \in \mcal{S} \rp \leq e^\varepsilon \Pr\lp M(D') \in \mcal{S} \rp+\delta,  $$
where $D = (X_1,...,X_n)$ and $D' = (X_1,...,X'_i,...,X_n)$ are neighboring pairs that can be obtained from each other by adding or removing all the records that belong to a particular user.
\end{definition}
In our frequency estimation setting (see Figure~\ref{fig:secagg_itform}), DP can be achieved in two different ways:
\begin{itemize}[topsep=0em, leftmargin=1.5em]
    \item Central-level DP criterion: $f(Y^n)$ is $(\varepsilon,\delta)$-DP.
    \item user-level DP criterion: $(Y_1,\dots, Y_n)$ is $(\varepsilon,\delta)$-DP.
\end{itemize}
The central DP criterion requires the server to apply a DP mechanism to its computation to obtain its final estimate $f(Y^n)$, and hence puts trust in the service provider. 
The user-level DP criterion removes the need for a trusted server as noise is added to each message before it is sent to the server.  
By the data processing property of DP, the latter is a stronger notion and implies the former.

In this section, we provide a secure and private frequency estimation scheme that satisfies the user-level DP criterion in addition to \eqref{S1} and \eqref{S2}. We characterize the accuracy-privacy trade-off achieved by this scheme as well as its per-user communication cost in Theorem~\ref{thm:private_secure_fe_achievability}. Since this scheme satisfies $(\varepsilon,\delta)$-user-level DP, it also satisfies the weaker $(\varepsilon,\delta)$-central DP criterion. Moreover, the accuracy-privacy trade-off achieved by this scheme is (nearly) optimal in the sense that it (nearly) matches the best trade-off achievable by any scheme satisfying the central DP criterion \citep{balcer2017differential}. Since our scheme is designed to satisfy the stronger user-level DP criterion, this means that we can achieve the optimal privacy-accuracy trade-off while removing the need for a trusted server. 
We show that the communication cost is also optimal by proving a lower bound on the communication cost of any scheme that achieves the optimal privacy-accuracy trade-off while satisfying \eqref{S1} and \eqref{S2}. 
In other words, any secure frequency estimation scheme requires at least as many bits to achieve the optimal privacy-accuracy trade-off. This establishes the optimality of our scheme in terms of communication cost.
Finally, we remark that although we present bounds in terms of standard DP (Definition~\ref{def:DP}), our scheme also satisfies R\'enyi differential privacy \citep{mironov2017renyi} (RDP), which allows for tighter privacy accounting when applying a private mechanism iteratively. 
We defer the details to Appendix~\ref{appendix:secure_private_fe}.

We next state the main results of this section starting with our achievability result.

\begin{theorem}[Private frequency estimation, informal] \label{thm:private_secure_fe_achievability}
The scheme presented in Section~\ref{subsec:private_and_secure} (see also Algorithm~\ref{alg1}) satisfies \eqref{S1}, \eqref{S2} and an $\lp \varepsilon, \delta \rp$-user-level DP criterion (and also $\lp \lp\alpha, \varepsilon/\log\lp \frac{1}{\delta} \rp \rp \rp$-RDP), while achieving
 \begin{itemize}[topsep=0em, leftmargin=1.5em]
     \item $\ell_\infty$ error $ \E\lb \lV \hat{\mu} - \mu\lp X^n \rp \rV_\infty\rb = O\lp \frac{\sqrt{\log d\log(1/\delta)}}{\varepsilon} \rp$;
     \item $\ell_2$ error $ \E\lb \lV \hat{\mu} - \mu\lp X^n \rp \rV^2_2\rb = O\lp \frac{n\log d\log(1/\delta)}{\varepsilon^2} \rp;$
 \end{itemize}     
 and uses $\tilde{O}\lp n \min\lp \varepsilon\sqrt{\log d/\log(1/\delta)}, \log d\rp \rp$ bits (where in $\tilde{O}$ we hide dependency on $\log n$ and $\log \log d$ terms).
\end{theorem}
The formal statement of Theorem~\ref{thm:private_secure_fe_achievability} and the proof are provided in Appendix~\ref{appendix:secure_private_fe} (see Theorem~\ref{thm:utility_guarantees}). Note that the $\lp \varepsilon, \delta \rp$-user-level DP guarantee in the theorem implies a $\lp \varepsilon, \delta \rp$-central DP guarantee. We contrast this with the optimal accuracy-privacy tradeoff achievable in the centralized case, i.e., when the only requirement imposed on the scheme is an $\lp \varepsilon, \delta \rp$-central DP criterion. 
For the $\ell_2$ and $\ell_\infty$ loss (i.e., setting the loss function in \eqref{C1prime} to be $\lV\cdot\rV_2$ or $\lV \cdot \rV_\infty$ respectively), the minimax error is well-known (see, for instance, \citep{hardt2010geometry, balcer2017differential}) as we state in the following lemma:
\begin{lemma}[Minimax error under central DP]\label{lemma:error_cdp}
    Under a $(\varepsilon, \delta)$-central DP, the minimax error  for frequency estimation, defined as
    $$ \min\limits_{M(\cdot) \text{ satisfies $(\eps,\delta)$-central DP}}\,\,\max\limits_{X^n}\,\, \E\lb \ell\lp M\lp X^n \rp, \mu\lp X^n \rp \rp\rb,$$
    is equal to
    \begin{itemize}[topsep=0em, leftmargin=1.5em]
        \item $\Theta\lp \frac{\min(\log d, \log(1/\delta)) }{\varepsilon} \rp$ under the $\ell_\infty$ loss; 
        \item $O\lp \frac{n\log d\log(1/\delta)}{\varepsilon^2} \rp$ under the $\ell^2_2$ loss; 
    \end{itemize}
\end{lemma}

We note that the the $\ell_\infty$ accuracy results in Theorem~\ref{thm:private_secure_fe_achievability}  matches that in Lemma~\ref{lemma:error_cdp} up to a $\max\lp \sqrt{\frac{\log d}{\log(1/\delta)}}, \sqrt{\frac{\log(1/\delta)}{\log d}} \rp$ factor, while the scheme in Theorem~\ref{thm:private_secure_fe_achievability} satisfies the additional \eqref{S1}, \eqref{S2} and the stronger user-level-DP constraints. This establishes the optimality of our scheme from an accuracy-privacy trade-off perspective. We also observe that the communication cost in Theorem~\ref{thm:private_secure_fe_achievability} \emph{decreases} with $\varepsilon$ when $\varepsilon \leq \log d$, meaning that we can compress more aggressively with more stringent privacy constraint. 
This behavior aligns with the conclusions of \cite{chen2022fundamental} (under a federated learning setting) and \cite{chen2020breaking} (under a local DP model). We next show that the communication cost  in Theorem~\ref{thm:private_secure_fe_achievability} is optimal under the $\ell_\infty$ loss (up to a $\msf{poly}\lp \log n, \log\log d \rp$ factor).   

\begin{corollary}\label{cor:private_secure_fe_lower_bound}
Any $(\varepsilon, \delta)$-central DP scheme that satisfies \eqref{S1} and \eqref{S2} such that:  \begin{itemize}[topsep=0em, leftmargin=1.5em]
     \item  $ \E\lb \lV \hat{\mu} - \mu\lp X^n \rp \rV_\infty\rb = O\lp \frac{\sqrt{\log d\log(1/\delta)}}{\varepsilon} \rp$ requires $\Omega\lp n \min\lp \varepsilon\sqrt{\log d / \log(1/\delta)}, \log d\rp\rp$ per-user communication;
     \item $ \E\lb \lV \hat{\mu} - \mu\lp X^n \rp \rV^2_2\rb = O\lp \frac{n\log d\log(1/\delta)}{\varepsilon^2} \rp $ requires $\Omega\lp n\min\lp\frac{\varepsilon}{\log(1/\delta)}, \log d\rp \rp$ per-user communication.
 \end{itemize} 
 \end{corollary}
 
 Recall from the previous section that we need $n \log d$ bits to securely compute the exact histogram. Corollary~\ref{cor:private_secure_fe_lower_bound} characterizes the reduction in communication cost when the histogram is computed approximately due to the privacy constraint and $\varepsilon = O\lp \log d \rp$.
 

\begin{lemma}\label{lemma:error_function}
    Let $R\lp \beta\rp$ be defined as in \eqref{eq:rate_function}. When $X_i \in \lbp e_1,...,e_d \rbp$ for $i \in [n]$, there is a worst-case prior distribution $\pi_{X^n}$ (possibly correlated for $X_i$'s), s.t.
    \begin{itemize}[topsep=0em, leftmargin=1.5em]
        \item under the $\ell_\infty$ loss , $R\lp \beta \rp = O\lp {n \log  d }/{\beta} \rp$;
        \item under the $\ell_2$ loss,  $R\lp \beta \rp = O\lp {n^2\log d}/{\beta} \rp$.
    \end{itemize}
\end{lemma}
Lemma~\ref{lemma:error_function} is a special case of Lemma~\ref{lemma:lossy_lower_bound}. However, to obtain the asymptotic scaling, we make use of Fano's inequality, with carefully constructed prior distributions via $\ell_\infty$ and $\ell_2$ packing over the space of all histograms. The proof can be found in Appendix~\ref{proof:error_function}


\subsection{Frequency Estimation via Noisy Sketch}\label{subsec:private_and_secure}
Next, we present a (nearly) optimal secure and private frequency estimation scheme in Algorithm~\ref{alg1} that uses the optimal communication in Corollary~\ref{cor:private_secure_fe_lower_bound}. We use the following  ingredients in our scheme: (1) the specific SecAgg implementation of \cite{bonawitz2016practical} (see Section~\ref{subsec:perfect_recovery_achievability} for a brief introduction), (2) count-sketch \citep{charikar2002finding} together with Hadamard transform, and (3) the Poisson-binomial mechanism \citep{chen2022poisson}.
Following the idea in Section~\ref{subsec:perfect_recovery_achievability}, we use the SecAgg protocol introduced by \cite{bonawitz2016practical} as a primitive and focus on designing $\lp \mcal{A}_{\msf{enc}}, \mcal{A}_{\msf{dec}}\rp$. Since $(\varepsilon, \delta)$-DP inevitably incurs $O\lp \frac{\log d}{\varepsilon} \rp$ error on the estimated frequency, it suffices to have SecAgg output an approximate sum (i.e., histogram) with distortion less than the DP error. 
This slack allows us to reduce the communication below the $\Omega\lp n\log d \rp$ lower bound per user for computing the exact histogram.

\textbf{Count-sketch.} We use count-sketch to achieve this goal. Count-sketch is a linear compression scheme (and hence can be represented in a matrix form $S = [S_1^\intercal, S_2^\intercal, ..., S_t^\intercal]\in \lbp -1, 0, 1 \rbp^{wt\times d}$ for some $w, t \in \mbb{N}$, where each $S_j \in \{-1, 0, 1\}^{w\times d}$ is generated according to an independent hash function) that allows for trading off  the estimation error for communication cost. A count-sketch is determined by two parameters $w, t\in\mbb{N}$; $w$ is the bucket size that controls the magnitude of $\ell_\infty$ error, and $t$, the number of hash functions, determines the failure probability. To apply count-sketch in the frequency estimation problem, each user computes a local sketch of its data, i.e., $SX_i$, and sends it to the server. Upon receiving local sketches, the server can unsketch and obtain an estimate on $\mu(X^n)$. By setting $t = \Theta\lp \log\lp {d}/{\gamma}\rp\rp$, count-sketch estimates $\mu$ with $O\lp {\lV \mu \rV_1}/{w}\rp$ error with failure probability at most $\gamma$\footnote{Here we apply an $\ell_1$ point-query bound due to the $\ell_1$ geometry of $\mu\lp X^n \rp$.}.

\textbf{Hadmard transform.}
After computing the local sketch, each user performs the Hadamard transform to flatten each $S_jX_i$ for $j \in [t]$ and $i \in [n]$, i.e., computes $H_wS_jX_i$, where $H_w$ is the (normalized) Walsh-Hadamard matrix (assuming $w$ is a power of $2$) satisfying the following relation:
$$ H_{2^n} = {\frac{1}{\sqrt{2}}\begin{bmatrix} {H_{2^{n-1}}}, &{H_{2^{n-1}}}\\{H_{2^{n-1}}}, &-{H_{2^{n-1}}}\end{bmatrix}}, \text{ and } H_0 = \begin{bmatrix}1 \end{bmatrix}.$$
The flattening step reduces the dynamic range of $S_iX_j$ in the sense that $\lV H_wS_iX_j \rV_\infty = \frac{1}{\sqrt{w}}\lV S_iX_j\rV_\infty$. This controls the $\ell_\infty$-sensitivity, which facilitates the following privatization steps.

\textbf{Poisson-binomial mechanism.}
Last, to introduce DP, we make use of the Poisson-binomial mechanism (PBM) \citep{chen2022poisson}. In PBM, users encode their locally flattened sketches $H_wSX_i$ into parameters of binomial random variables (and hence the sum of $n$ users' noisy reports follow a Poisson-binomial distribution). 
The main advantages of using PBM include: (1) the binomial distribution is closed under addition, and hence it is compatible with SecAgg;  (2) it asymptotically converges to a Gaussian distribution and gives R\'enyi DP guarantees (which supports tight privacy accounting); (3) it does not require modular clipping and hence results in an unbiased estimate of $\mu$ (as opposed to other user-level discrete DP mechanisms, such as those of \citet{ kairouz2021distributed, agarwal2021skellam}).

\begin{algorithm}[t]
	\SetAlgoLined
	\KwIn{users' data (one-hot) $X_1,...,X_n$, sketch parameters $w, t$, failure probability $\gamma$, PBM parameters $L, \theta$.}
	\KwOut{frequency estimate $\hat{\mu}$}
	Server broadcasts $t$ i.i.d. generated sketch matrices $S_1,...,S_t\in\lbp -1, 0, 1 \rbp^{w\times d}$\;
	\For{user $i \in [n]$}{
	Compute $t$ sketches $S_1X_i,...,S_tX_i$\;
	Perform Hadamard transform on each sketch\;
	Apply PBM on each transformed sketch\;
	}
    Server aggregates local noisy sketches via SecAgg, decodes PBM, and applies inverse Hadmard transform to obtain a noisy estimate $\widehat{S\mu}$\;
    Server unsketches $\widehat{S\mu}$ and obtains $\hat{\mu}$\;
	\KwRet{$\hat{\mu}$}
	\caption{Secure and private frequency estimation}\label{alg1}
\end{algorithm}

By putting these pieces together, we arrive at Algorithm~\ref{alg1} with privacy guarantees, estimation error, and communication cost as stated in Theorem~\ref{thm:private_secure_fe_achievability}. A more detailed version is given in Algorithm~\ref{alg2} in Appendix~\ref{appendix:secure_private_fe}. In addition, in Appendix~\ref{appendix:sparse_fe}, we show that we can improve the accuracy when additional knowledge on the sparsity of $\mu(X^n)$ is available.

Comparing the communication cost in Theorem~\ref{thm:private_secure_fe_achievability} and the lower bounds in Corollary~\ref{cor:private_secure_fe_lower_bound}, we see that under $\ell_\infty$ loss, Algorithm~\ref{alg1} matches the lower bound up to a $\log n$ and $\sqrt{\log d}$ factor, where the small sub-optimality gap is due to the modular arithmetic used by SecAgg. Closing this gap is left as a future work.

\section{Experiments}\label{sec:experiments}
In this section, we provide empirical results for Algorithm~\ref{alg1}, which we label as `Sketched PBM' .

We compare sketched PBM with other decentralized (local) DP mechanisms, including randomized response (RR) \citep{warner1965randomized, kairouz16} and the Hadamard response (HR) \citep{acharya2019hadamard} (which is order-wise optimal for all $\varepsilon = O\lp \log d \rp$)

\begin{figure}
    \centering
         \includegraphics[width=0.44\textwidth]{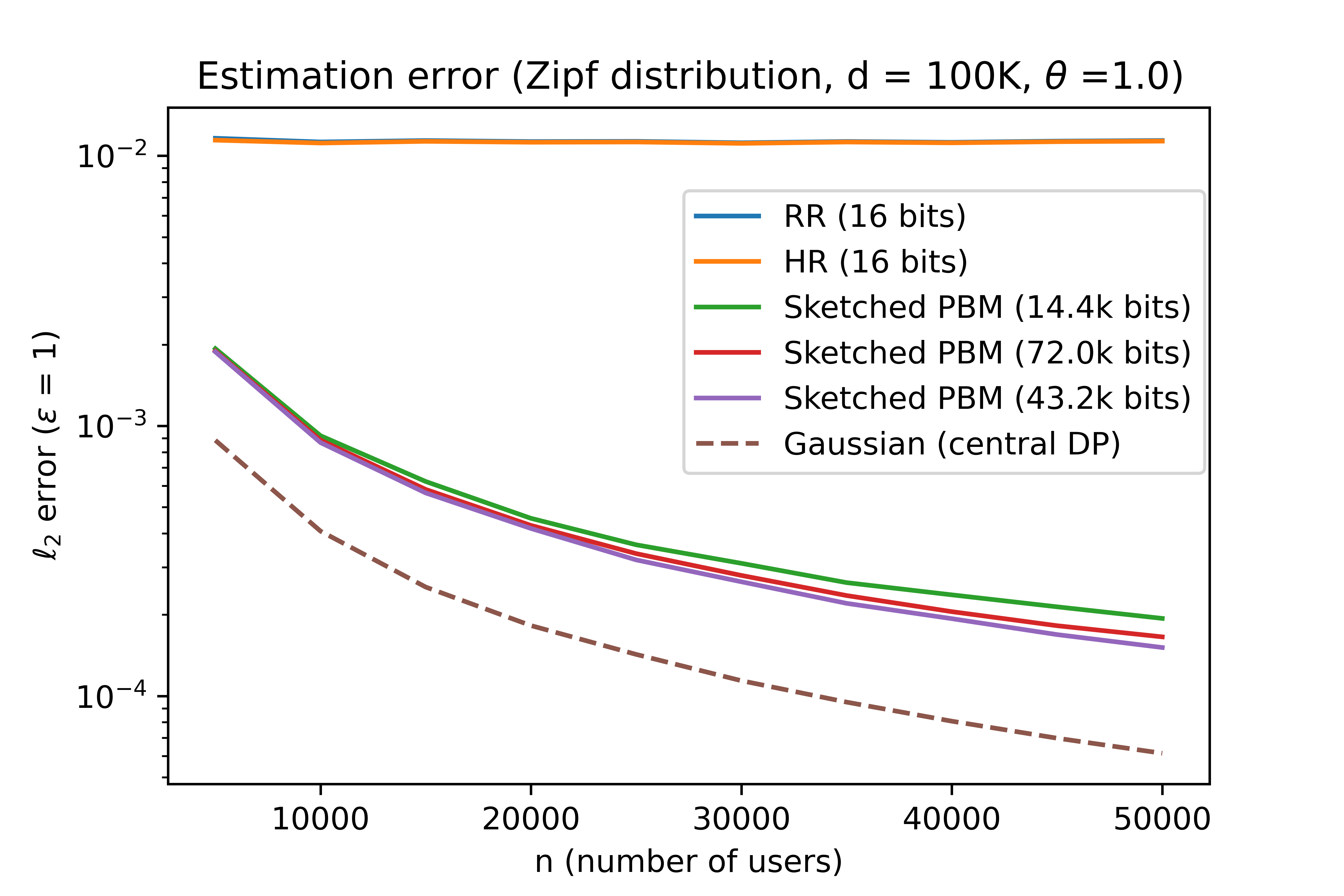}
     \caption{$\ell_2$ loss with $\varepsilon = 1$. The error is computed with a normalization (the goal is to estimate $\frac{\mu\lp X^n \rp}{n}$). 
     }\label{fig2}
\end{figure}

We set $d = 10^5$ and $n \in [10K, 50K]$, i.e., in a regime where $d \gg n$. Under this regime, it is well-known that local DP suffers from poor-utility \citep{duchi2013local}. We demonstrate that our proposed sketched PBM achieves a much better convergence rate (though admittedly at the cost of higher communication as predicted by our theoretical results). We also remark that the communication cost per user of the sketched PBM is fixed in this set of experiments, and thus the (normalized) estimation error does not strictly decrease with $n$ (recall that our theory suggests in order to achieve the best performance, the communication cost has to be \emph{increasing with} $n$). More detailed empirical results can be found in Appendix~\ref{appendix:experiments}.

\newpage
\bibliography{references}

\onecolumn
\appendix
\section{Additional Details of Section~\ref{sec:secure_private_fe}}\label{appendix:secure_private_fe}
In this section, we provide additional details and empirical results of our noisy sketch scheme Algorithm~\ref{alg1} in Section~\ref{sec:secure_private_fe} and give formal proofs on its privacy and utility guarantees. As mentioned in Section~\ref{sec:secure_private_fe} our goal is to design a scheme that satisfies a stronger version of (distributed) DP, i.e., R\'enyi differential privacy, as it allows for tight privacy accounting. Therefore, in this section we  first provide an RDP guarantee for our scheme, and then convert the RDP guarantee to $(\varepsilon, \delta)$-DP using well-known conversion results such as \cite{mironov2017renyi}. To this end, we start by giving a brief introduction to R\'enyi DP.

\subsection{R\'enyi Differential Privacy (RDP)}
A useful variant of DP is the R\'enyi differential privacy (RDP), which allows for tight privacy accounting when a mechanism $M$ is applied iteratively. 

\begin{definition}[R\'enyi Differential Privacy (RDP)]\label{def:RDP}
A randomized mechanism $M$ satisfies $(\alpha, \varepsilon)$-RDP if for any two neighboring datasets $D, D'$, we have that $D_\alpha\lp P_{M(D)}, P_{M(D')} \rp\leq \varepsilon$ where $D_{\alpha}\lp P, Q\rp$ is the  R\'enyi divergence between $P$ and $Q$ and is given by 
$$ D_\alpha\lp P, Q \rp \eqDef \frac{1}{\alpha}\log\lp \E_{Q}\lb \lp \frac{P(X)}{Q(X)} \rp^\alpha \rb \rp.$$
\end{definition}

Note that one can convert an RDP guarantee to an (approximate) DP guarantee (for instance, see \cite{mironov2017renyi}) but not the other way around in general. Although we presented our bounds in Section~\ref{sec:secure_private_fe} in terms  of approximate DP, our proposed schemes  satisfy the RDP definition as we show next.

\subsection{Details of Algorithm~\ref{alg1}}
We start by briefly recalling the details of count-sketch \cite{charikar2002finding}, which serves as our main compression tool for reducing communication costs. Count-sketch can be constructed via two sets of (pairwise independent) hash functions $h_i:[d] \ra [w]$ and $\sigma_i:[d]\ra\lbp -1, +1\rbp$ for $i \in [t]$. The functions can be organized in matrix form $S \in \lbp -1, 0, 1 \rbp^{wt\times d}$, which can be viewed as a vertical stack of $S_1,...,S_t \in \{ -1, 0, 1\}^{w\times d}$, where for $i \in [t]$, $ \lp S_i\rp_{j, k} = \sigma_i(j)\cdot\bbm{1}_{\lbp h_i(j)=k\rbp}$. Note that $m \eqDef w\cdot t$ is the embedded dimension.

In Algorithm~\ref{alg2}, we give a more detailed description of Algorithm~\ref{alg1}, our private frequency estimation scheme from Section~\ref{sec:secure_private_fe}. We analyze the performance of Algorithm~\ref{alg2} in the next section.

\subsection{Performance Analysis for Algorithm~\ref{alg2}}

\begin{algorithm}
	\SetAlgoLined
	\KwIn{users' data $ X_1,...,X_n \subseteq \lbp e_1,...,e_d \rbp$, failure probability $\gamma$, sketch parameter $w, t$, PBM parameter $L, \theta$}
	\KwOut{frequency estimate $\hat{\mu}$}
	Server generates $(S_1,...,S_t)$ (with $t = \Theta\lp \log\lp \frac{d}{\gamma} \rp \rp$ and $w$ being a power of two and satisfying $w = \Theta\lp \min\lp n, \frac{n\varepsilon}{t}\rp \rp$)\;
	The server broadcasts $S_1,...,S_t$ to all users\;
	\For{$i \in [n]$}{
	Set $\theta = $ and $L = $\;
	\For {$j \in [t]$}{
	user $i$ computes $p_{ij} = \theta\lp H_w\cdot S_j\cdot X_i\rp + 1/2$, where $H_w \in \{-1/\sqrt{w}, 1/\sqrt{w}\}^{w \times w}$ is the Hadamard matrix\;
	user $i$ generates $Y_{ij} \eqDef \msf{Binom}\lp L, p_{ij} \rp$ coordinate-wisely (so $Y_{ij} \in [L]^{w}$)\;
	}
	}
    The server aggregates (via SecAgg \cite{bonawitz2016practical}) noisy reports $\{ Y_{ij }\}$ and computes the median
    $$ \lp \hat{S_1\mu}, ..., \hat{S_t\mu} \rp \eqDef \lp \frac{1}{\theta\sqrt{w}}\sum_{i=1}^n H_w\cdot\lp \frac{Y_{i1}}{L} -\frac{1}{2} \rp,..., \frac{1}{\theta\sqrt{w}}\sum_{i=1}^n H_w\cdot\lp\frac{Y_{it}}{L} -\frac{1}{2} \rp \rp. $$
    Server unsketches by computing the median:
    $$ \hat{\mu} = \msf{median}\lp S_1^\intercal\hat{S_1\mu}, ..., S_t^\intercal\hat{S_t\mu} \rp. $$
	\KwRet{$\hat{\mu}$}
	\caption{Secure and private frequency estimation with noisy sketch (detailed)}\label{alg2}
\end{algorithm}

We start by proving that Algorithm~\ref{alg2} satisfies the following RDP guarantee.
\begin{theorem}[RDP guarantee]\label{thm:rdp}
As long as $\theta \leq \frac{1}{4}$, Algorithm~\ref{alg2} satisfies $(\alpha, \tau(\alpha))$-RDP for all $\alpha > 1$ and $\tau(\alpha)$ such that
$$ \tau(\alpha) \geq C_0 \frac{\theta^2L \alpha}{n}\cdot wt,$$
for some $C_0 > 0$
\end{theorem}
\textbf{Proof.}
The proof follows from \citep[Corollary~3.2]{chen2022poisson}.
\hfill$\blacksquare$

Once we obtain an RDP guarantee, we cast it into an $(\varepsilon, \delta)$-DP guarantee by using results due to \cite{canonne2020discrete}.
\begin{theorem}[$(\varepsilon, \delta)$-DP guarantee]\label{thm:approx_dp}
Assume $\delta \leq \exp\lp -\frac{m\theta^2wt}{n} \rp$. Then Algorithm~\ref{alg2} satisfies an $(\varepsilon, \delta)$ distributed DP guarantee for all $\varepsilon$ and $\delta$ satisfying
$$ \varepsilon = \Omega\lp \sqrt{\frac{L\theta^2wt\log\lp \frac{1}{\delta} \rp}{n}} \rp. $$
\end{theorem}
\textbf{Proof.}
We apply \citep{canonne2020discrete} to convert the RDP guarantee in Theorem~\ref{thm:rdp}.
\begin{lemma}[Renyi DP to approximate DP] \label{lemma:rdp_to_approx_dp}
For any $\alpha \in (1, \infty)$, if $$ D_\alpha\lp \mcal{M}\lp \bm{x}\rp \middle\Vert \mcal{M}\lp \bm{x}' \rp \rp \leq \tau, $$
then  $\mcal{M}(\cdot)$ satisfies $(\varepsilon, \delta)$-DP for 
$$ \varepsilon \geq \varepsilon^*\eqDef \tau+\frac{\log\lp \frac{1}{\delta} \rp+(\alpha-1)\log\lp1-\frac{1}{\alpha}\rp-\log\lp \alpha \rp}{\alpha-1}. $$
\end{lemma}

Applying Theorem~\ref{thm:rdp} and Lemma~\ref{lemma:rdp_to_approx_dp} above and plugging in $\tau = C_0 \frac{\theta^2L \alpha}{n}\cdot wt$, we see that $\hat{\mu}$ is $(\varepsilon, \delta)$-DP for
\begin{align*}
    \varepsilon^* &= C_0 \frac{\theta^2L \alpha}{n}\cdot wt+\frac{\log\lp \frac{1}{\delta} \rp+(\alpha-1)\log\lp1-\frac{1}{\alpha}\rp-\log\lp \alpha \rp}{\alpha-1}\\
    &\leq C_0 \frac{\theta^2L}{n}\cdot wt + C_0 \frac{\theta^2L (\alpha-1)}{n}\cdot wt+\frac{\log\lp \frac{1}{\delta} \rp}{\alpha-1}\\
    &\overset{\text{(a)}}{=}C_0 \frac{\theta^2L}{n}\cdot wt+2\sqrt{C_0\frac{L\theta^2wt\log\lp \frac{1}{\delta} \rp}{n}}\\
    &\overset{\text{(b)}}{=} O\lp \sqrt{\frac{L\theta^2wt\log\lp \frac{1}{\delta} \rp}{n}} \rp,
\end{align*}
where (a) holds if we pick $\alpha-1 = \sqrt{\frac{n\log(1/\delta)}{\theta^2Lwt}}$ (i.e., such that AM-GM inequality holds with equality), and (b) holds if 
$$\log\lp \frac{1}{\delta}\rp \geq \frac{L\theta^2wt}{n} \Longleftrightarrow \delta \leq \exp\lp -\frac{L\theta^2wt}{n} \rp.$$ 
\hfill$\blacksquare$


Finally, in the following theorem, we compute the communication cost and control the $\ell_\infty$ and $\ell_2$ estimation error of our algorithm.

\begin{theorem}[Privacy and Utility of Algorithm~\ref{alg2}]\label{thm:utility_guarantees}
Let 
\begin{align*}
    &t = \log\lp \frac{d}{\gamma} \rp,\\
    &w= \min\lp n,\lp \frac{n\varepsilon}{\sqrt{\log\lp \frac{d}{\gamma}\rp\log\lp \frac{1}{\delta} \rp}} \rp\rp,\\
    &L = \left\lceil \frac{n\varepsilon^2}{wt\log(1/\delta)} \right\rceil+1,\\
    &\theta = O\lp \min\lp \frac{1}{4}, \sqrt{\frac{n\varepsilon^2}{wt\log(1/\delta)}} \rp \rp.
\end{align*}
Let $\hat{\mu}$ be the output of the Algorithm~\ref{alg2}.  Then:
\begin{itemize}[topsep=0em, leftmargin=1.5em]
    \item $\hat{\mu}$ satisfies $(O(\varepsilon), \delta)$-DP and $\lp \alpha, O\lp \frac{\varepsilon^2\alpha}{\log(1/\delta)} \rp \rp$-R\'enyi DP.
    \item The communication complexity is $\tilde{O}(\min \lp n\varepsilon\log\lp\frac{1}{\gamma}\rp, n\rp)$ bits per user. 
    \item With probability at least $1-\gamma$,
$$ \lV \hat{\mu} - \mu\rV_\infty = \max_{j\in[d]}\lba \mu_j - \hat{\mu}_j \rba \leq \frac{4n}{w}+O\lp\frac{\sqrt{\log\lp \frac{d}{\gamma}\rp\log\lp \frac{1}{\delta} \rp}}{\varepsilon}\rp = O\lp \frac{\sqrt{\log\lp \frac{d}{\gamma}\rp\log\lp \frac{1}{\delta} \rp}}{\varepsilon} \rp.$$
\item By setting $\hat{\mu}_j = 0$ for all $j\in[d]$ such that $\hat{\mu}_j = O\lp \frac{\log\lp \frac{d}{\gamma} \rp\log \frac{1}{\delta}}{\varepsilon} \rp$, the $\ell^2_2$ estimation error is bounded by $$ O\lp\frac{n\log^2\lp \frac{d}{\gamma} \rp\log\lp \frac{1}{\delta} \rp}{\varepsilon^2}\rp.$$
\end{itemize}
\end{theorem}

\textbf{Proof.}
\paragraph{Privacy guarantee.}
By plugging $L = \left\lceil \frac{n\varepsilon^2}{wt\log(1/\delta)} \right\rceil+1$ and $\theta = O\lp \min\lp \frac{1}{4}, \sqrt{\frac{n\varepsilon^2}{wt\log(1/\delta)}} \rp \rp$ into Theorem~\ref{thm:rdp} and Theorem~\ref{thm:approx_dp}, we immediately obtain the desired privacy guarantee.

\paragraph{Analysis of the communication cost.} 
Let $\mbb{Z}^m_M$ be the finite group that SecAgg operates on. In Algorithm~\ref{alg2}, client $i$ needs to communicate $\lbp Y_{ij} | i = 1,...,t \rbp$ to the server. Notice that each $Y_{ij} \in [L]^w$, but for all $j \in [t]$, each coordinate of $\sum_i Y_{ij}$ can be as large as $nL$. Therefore, we will set $M = nL$. Now, if $w = \lp \frac{n\varepsilon}{\sqrt{\log\lp \frac{d}{\gamma}\rp\log\lp \frac{1}{\delta} \rp}} \rp \leq n$, then the communication cost for each client becomes 
\begin{align*}
     &m\log(M+1) = m\log(nL+1) = wt\log(nL+1) \\
     & = \frac{n\varepsilon\sqrt{\log(d/\gamma)}}{\sqrt{\log\lp1/\delta\rp}}\log\lp n\lp \left\lceil \frac{n\varepsilon^2}{wt\log(1/\delta)} \right\rceil+1 \rp+1 \rp\\
     & = \frac{n\varepsilon\sqrt{\log(d/\gamma)}}{\sqrt{\log\lp1/\delta\rp}} \log\lp n \lp\left\lceil \frac{\varepsilon}{\sqrt{\log(d/\gamma)\log(1/\delta)}} \right\rceil+1\rp+1\rp\\
     & = \tilde{O}\lp \frac{n\varepsilon\sqrt{\log(d/\gamma)}}{\sqrt{\log(1/\delta)}} \rp,
\end{align*}
where in the last equation we hide the $\log(n\lceil \varepsilon \rceil)$ term into $\tilde{O}(\cdot)$ for simplicity. On the other hand, if $w = n$, then
\begin{align*}
     &m\log(M+1) =wt\log(nL+1) \\
     & = n\log\lp \frac{d}{\gamma} \rp\log\lp n\lp \left\lceil \frac{n\varepsilon^2}{wt\log(1/\delta)} \right\rceil+1 \rp+1 \rp\\
     & = n\log\lp \frac{d}{\gamma} \rp\log\lp n\lp \left\lceil \frac{\varepsilon^2}{ \log\lp \frac{d}{\gamma} \rp\log(1/\delta)} \right\rceil+1 \rp+1 \rp\\
     & = \tilde{O}\lp n\log(d/\gamma) \rp.
\end{align*}
    
\paragraph{Bounding the $\ell_\infty$ error.}  
We apply a similar analysis of error bounds using the count-sketch. 
Let $$ \hat{\mu}^{(j)} = S_j^\intercal\hat{S_j\mu} = S_j^\intercal \lp \frac{1}{\theta\sqrt{w}}\sum_{i=1}^nH_w\cdot\lp \frac{Y_{ij}}{L}-\frac{1}{2} \rp \rp,$$ for $j \in [t]$.
Define $N^{(j)} \in \mbb{R}^w$ be the estimation error of the $j$-th sketch, i.e.,
$$ N^{(j)} \eqDef  S_j \mu -  \frac{1}{\theta\sqrt{w}}\sum_{i=1}^nH_w\cdot\lp \frac{Y_{ij}}{m}-\frac{1}{2} \rp. $$
Then, for any $i \in [d]$, we can write the absolute error of the $j$-th sketch as
$$ \hat{\mu}^{(k)}_i - \mu_i = \sum_{j\neq i} \sigma_k(j)\sigma_k(i)\bbm{1}_{\lbp h(j) = h(i) \rbp}\mu_{j}+  N^{(j)}_{h_k(i)}.  $$

Therefore, we must have
\begin{align*}
    \E\lb \lba \hat{\mu}^{(k)}_i - \mu_i\rba\rb 
    & \leq \E\lb \lba\sum_{j\neq i} \sigma_k(j)\sigma_k(i)\bbm{1}_{\lbp h_k(j) = h_k(i) \rbp}\mu_{j}\rba+ \lba  N^{(j)}_{h_k(i)}\rba\rb \\
    & \overset{\text{(a)}}{\leq} \E\lb \sum_{j\neq i} \bbm{1}_{\lbp h_k(j) = h_k(i) \rbp}\mu_{j}\rb + \sqrt{\E\lb  \lp N^{(j)}_{h_k(i)}\rp^2\rb} \\
    & \overset{\text{(b)}}{\leq} \frac{n}{w} + \sqrt{\E\lb \lp N^{(j)}_{h_k(i)}\rp^2\rb},
\end{align*}
where (a) follows due to Jensen's inequality and the fact that $\sigma_k(\cdot) \in \lbp -1, +1 \rbp$, (b) holds since $h_k(i)$ and $h_k(j)$ are pairwise independent. 

Next, we upper bound $\E\lb \lp N^{(j)}_{h_k(i)}\rp^2\rb$. For notational simplicity, assume $h_k(i) = h \in [w]$. Observe that 
\begin{align*}
    N^{(j)} & = S_j \mu -  \frac{1}{\theta\sqrt{w}}\sum_{i=1}^nH_w\cdot\lp \frac{Y_{ij}}{L}-\frac{1}{2} \rp \\
    & = H_w\lp H_wS_j \mu - \frac{1}{\theta\sqrt{w}}\sum_{i=1}^n\lp \frac{Y_{ij}}{L}-\frac{1}{2} \rp\rp,
\end{align*}
where the second equality is due to the fact that $H_w\cdot H_w = I_w$. 

Denote 
$$ \Delta_j \eqDef H_wS_j \mu - \frac{1}{\theta\sqrt{w}}\sum_{i=1}^n\lp \frac{Y_{ij}}{L}-\frac{1}{2} \rp \in \mbb{R}^w. $$
Note that $H_wS_j \mu$ is the input to the PBM and $\frac{1}{\theta\sqrt{w}}\sum_{i=1}^n\lp \frac{Y_{ij}}{L}-\frac{1}{2}\rp$ is the estimate of PBM, so $\Delta_j$ satisfies the following properties (see \citep{chen2022poisson} for more details):
\begin{itemize}
    \item $\Delta_j(h)$ is independent of $\Delta_j(h')$ for all $h \neq h'$ (where $\Delta_j(h)$ is the $h$-th coordinate of $\Delta_j$).
    \item $\E\lb \Delta_j \rb = 0$.
    \item For any $h \in [w]$, $\E\lb\Delta^2_j(h)\rb = \frac{1}{wL\theta^2}\sum_{i=1}^n \Var\lp Y_{ij} \rp \leq \frac{n}{4wL\theta^2}$.
\end{itemize}
Let $H_w(h)$ be the $h$-th row of $H_w$. Then
\begin{align*}
    \E\lb\lp N^{(j)}_h\rp^2\rb 
    = \E\lb \left\lan H_w(h),  \Delta_j\right\ran^2 \rb 
     \overset{\text{(a)}}{=} \E\lb \frac{1}{w}\lV\Delta_j\rV^2 \rb
     \overset{\text{(b)}}{\leq} \frac{n}{4wL\theta^2} 
     \overset{\text{(c)}}{=} \frac{n}{4w}O\lp\frac{wt\log(1/\delta)}{n\varepsilon^2}\rp 
     = O\lp\frac{t\log(1/\delta)}{\varepsilon}\rp,
\end{align*}
where (a) holds since each coordinate of $\Delta_j$ is independent and each coordinate of $H_w(h)$ is either $\frac{1}{\sqrt{w}}$ or $-\frac{1}{\sqrt{w}}$, (b) holds since $\E\lb\Delta^2_j(h)\rb \leq \frac{n}{4wL\theta^2}$ for all $h \in [w]$, and (c) is because of our choice of $L$ and $\theta$.


Therefore, by Markov's inequality, we have
$$ \Pr\lbp \lba \hat{\mu}^{(k)}_i - \mu_i\rba \geq \frac{4n}{w} + O\lp\frac{\sqrt{t\log\lp \frac{1}{\delta} \rp}}{\varepsilon}\rp\rbp \leq \frac{1}{4}. $$
Taking the median for $\lp \hat{\mu}^{(1)}_i, ..., \hat{\mu}^{(t)}_i\rp$ to apply the Chernoff bound, we obtain 
\begin{align*}
    \Pr\lbp \lba \hat{\mu}_i - \mu_i\rba \geq  \frac{4n}{w} + O\lp\frac{\sqrt{t\log\lp \frac{1}{\delta} \rp}}{\varepsilon}\rp\rbp 
    & \leq \Pr\lbp\sum_{k=1}^t \bbm{1}_{\lbp \lba \hat{\mu}^{(k)}_i - \mu_i\rba \geq \frac{4n}{w} + O\lp\frac{\sqrt{t\log\lp \frac{1}{\delta} \rp}}{\varepsilon}\rp \rbp}  \geq \frac{t}{2}\rbp\\ 
    & \leq \Pr\lbp\msf{Binom}\lp t, \frac{1}{4} \rp \geq \frac{t}{2}\rbp\\
    & \leq \frac{\gamma}{d},
\end{align*}
if we take $t = O\lp\log\lp\frac{d}{\gamma}\rp\rp$, where the last inequality is due to the Chernoff bound.

Taking the union bound over $i \in [d]$, we conclude that 
$$ \Pr\lbp \max_{i \in [d]}\lba \hat{\mu}_i - \mu_i\rba \geq \frac{4n}{w} + O\lp\frac{\sqrt{t\log\lp \frac{1}{\delta} \rp}}{\varepsilon}\rp\rbp \leq \gamma. $$

Setting $w =  O\lp \frac{n\varepsilon}{ \sqrt{t\log\lp \frac{1}{\delta} \rp}} \rp = O\lp \frac{n\varepsilon}{ \sqrt{\log\lp \frac{d}{\gamma}\rp\log\lp \frac{1}{\delta} \rp}} \rp$, we arrive at the desired result.

\paragraph{Bounding the $\ell_2$ error.}
Since
$$ \Pr\lbp \max_{j \in [d]}\lba \hat{\mu}_i - \mu_i\rba = O\lp\frac{\sqrt{t\log\lp \frac{1}{\delta} \rp}}{\varepsilon}\rp\rbp \leq \gamma,  $$  
we condition on the event 
$$\mcal{E} \eqDef \lbp \max_{j \in [d]}\lba \hat{\mu}_i - \mu_i\rba = O\lp\frac{\sqrt{t\log\lp \frac{1}{\delta} \rp}}{\varepsilon}\rp\rbp. $$
Under $\mcal{E}$, when thresholding out every coordinate $i$ such that $\hat{\mu}_i \leq O\lp\frac{\sqrt{t\log\lp \frac{1}{\delta} \rp}}{\varepsilon}\rp$ (denoted as $\check{\mu}_i$), we must have
$$
\begin{cases}
\lba \check{\mu}_i - \mu_i\rba \leq O\lp\frac{\sqrt{t\log\lp \frac{1}{\delta} \rp}}{\varepsilon}\rp, \text{ if } \mu_i \neq 0 \\
\lba \check{\mu}_i - \mu_i\rba = 0, \text{ if }\mu_i = 0.
\end{cases}
$$
Since there can be at most $n$ coordinates such that $\mu_i \neq 0$, the $\ell^2_2$ error can be at most
$$ \sum_{i=1}^d \lp \check{\mu}_i - \mu_i \rp^2 \leq  n\cdot\frac{t\log\lp \frac{1}{\delta} \rp}{\varepsilon^2}+(d-n)\cdot 0 =  O\lp\frac{n\log\lp \frac{d}{\gamma} \rp\log\lp \frac{1}{\delta} \rp}{\varepsilon^2}\rp. $$
This completes the proof of Theorem~\ref{thm:utility_guarantees}.

Finally, setting $\gamma = \frac{1}{\msf{poly}\lp n, d \rp}$, we can cast the high-probability bound in Theorem~\ref{thm:utility_guarantees} into expected bounds shown in Theorem~\ref{thm:private_secure_fe_achievability}.

\section{Additional Experiments}\label{appendix:experiments}

\begin{figure}
    \centering
         \includegraphics[width=0.95\textwidth]{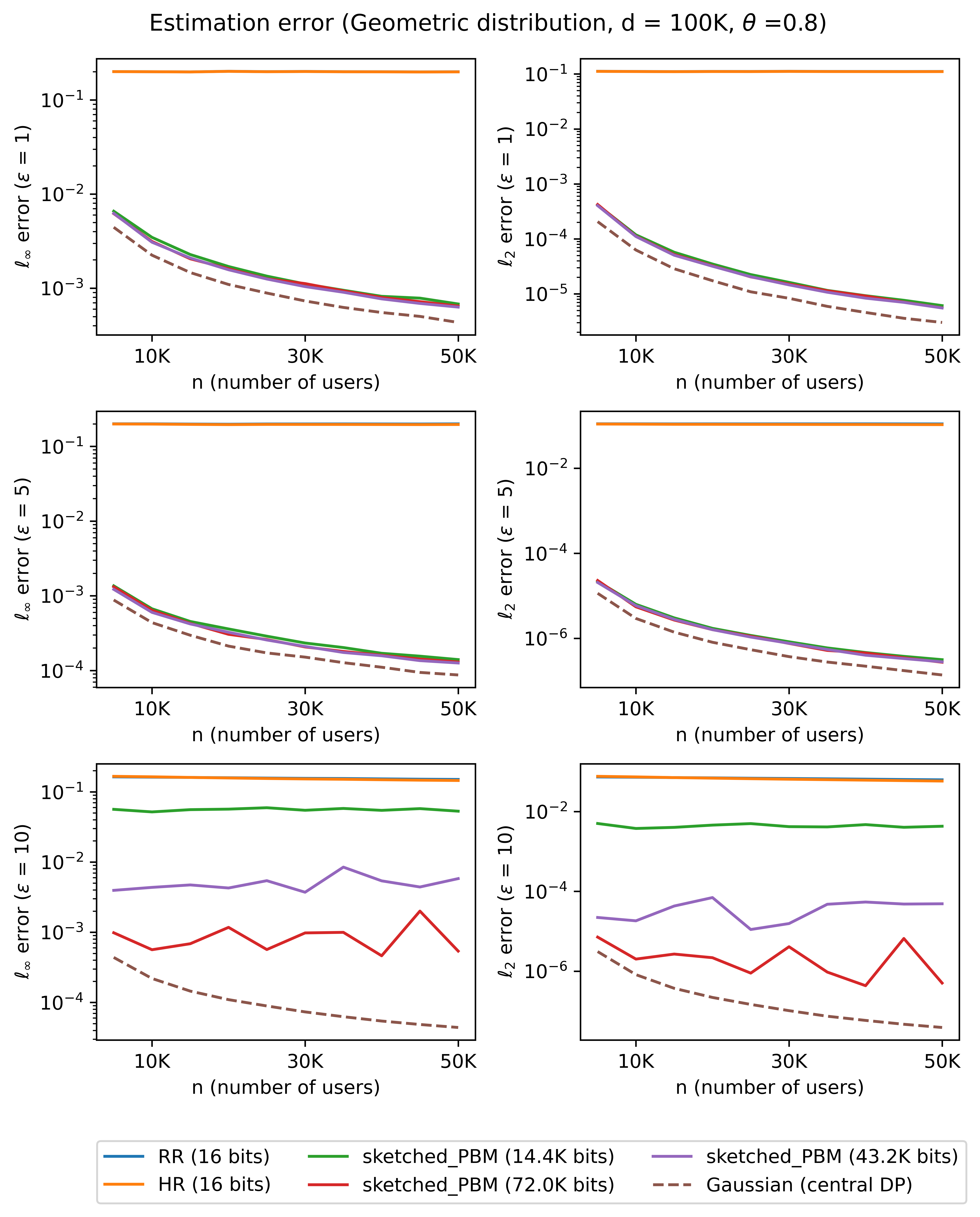}
     \caption{$\ell_\infty$ and $\ell_2$ loss with $\varepsilon = \{1, 5, 10\}$. The error is computed with a normalization (the histogram is normalized by a factor of $n$, i.e., $\frac{\mu\lp X^n \rp}{n}$). The $y$-axis is under a log-scale. In addition, when computing the $\ell_2$ error, we project all the estimated histograms into the probability simplex to further reduce the estimation error (also been adopted by \cite{acharya2019hadamard}).}\label{fig1}
\end{figure}

In this section, we provide additional empirical results for Algorithm~\ref{alg1}, which we label as `sketched PBM'. 
%
As in Section~\ref{sec:experiments}, in the first set of experiments, we compare sketched PBM with other decentralized (local) DP mechanisms, including randomized response (RR) \citep{warner1965randomized, kairouz16} and the Hadamard response (HR) \citep{acharya2019hadamard} (which is order-wise optimal for all $\varepsilon = O\lp \log d \rp$)\footnote{For the local DP mechanisms, we partly use the implementation from \url{https://github.com/zitengsun/hadamard_response}.}. The data is generated under a (truncated) Geometric distribution (with $\theta = 0.8$) in Figure~\ref{fig1} and under a (truncated) Zipf distribution (with $\theta = 1.0$) in Figure~\ref{fig2}. For the (centralized) Gaussian and the (distributed) sketched PBM mechanisms, $\delta$ is set to be $10^{-5}$. For sketched PBM, we set the parameter $L=10$.

We set $d = 10^5$  and $n \in [10k, 50k]$, i.e., in a regime where $d \gg n$ and compare the above schemes for $\varepsilon \in \{1, 5, 10\}$. Under this regime, it is well-known that local DP suffers from poor-utility \citep{duchi2013local}. We demonstrate that our proposed sketched PBM mechanism achieves a much better convergence rate (though admittedly at the cost of higher communication) both for the Gemoetric and Zipf distributions. We also remark that the per user communication cost  of the sketched PBM mechanism is fixed in this set of experiments, and thus the (normalized) estimation error does not strictly decrease with $n$ (recall that our theory suggests in order to achieve the best performance, the per user communication cost has to be \emph{increasing with} $n$). We note that in the low privacy regimes (e.g., when $\varepsilon = 10$), the communication budget has a greater impact on the accuracy of sketched PBM. This suggests that in this regime the performance of the scheme is limited by the compression error. Equivalently, the number of bits used by the scheme are below the threshold characterized by our theory to achieve the central DP performance.


\begin{figure}
    \centering
         \includegraphics[width=0.95\textwidth]{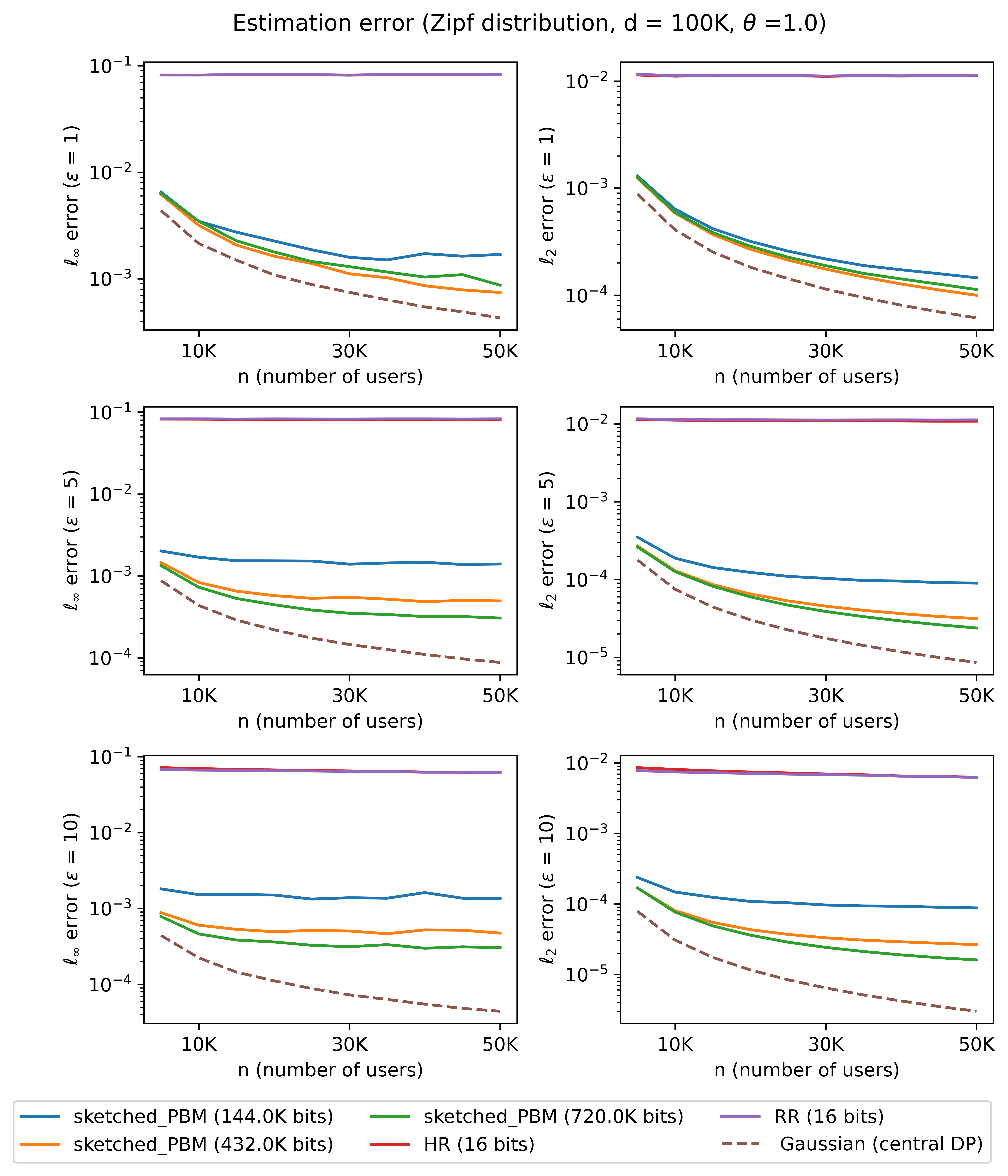}
     \caption{$\ell_\infty$ and $\ell_2$ loss with $\varepsilon = \{1, 5, 10\}$.}\label{fig2}
\end{figure}

\newpage

\begin{figure}
    \centering
         \includegraphics[width=0.95\textwidth]{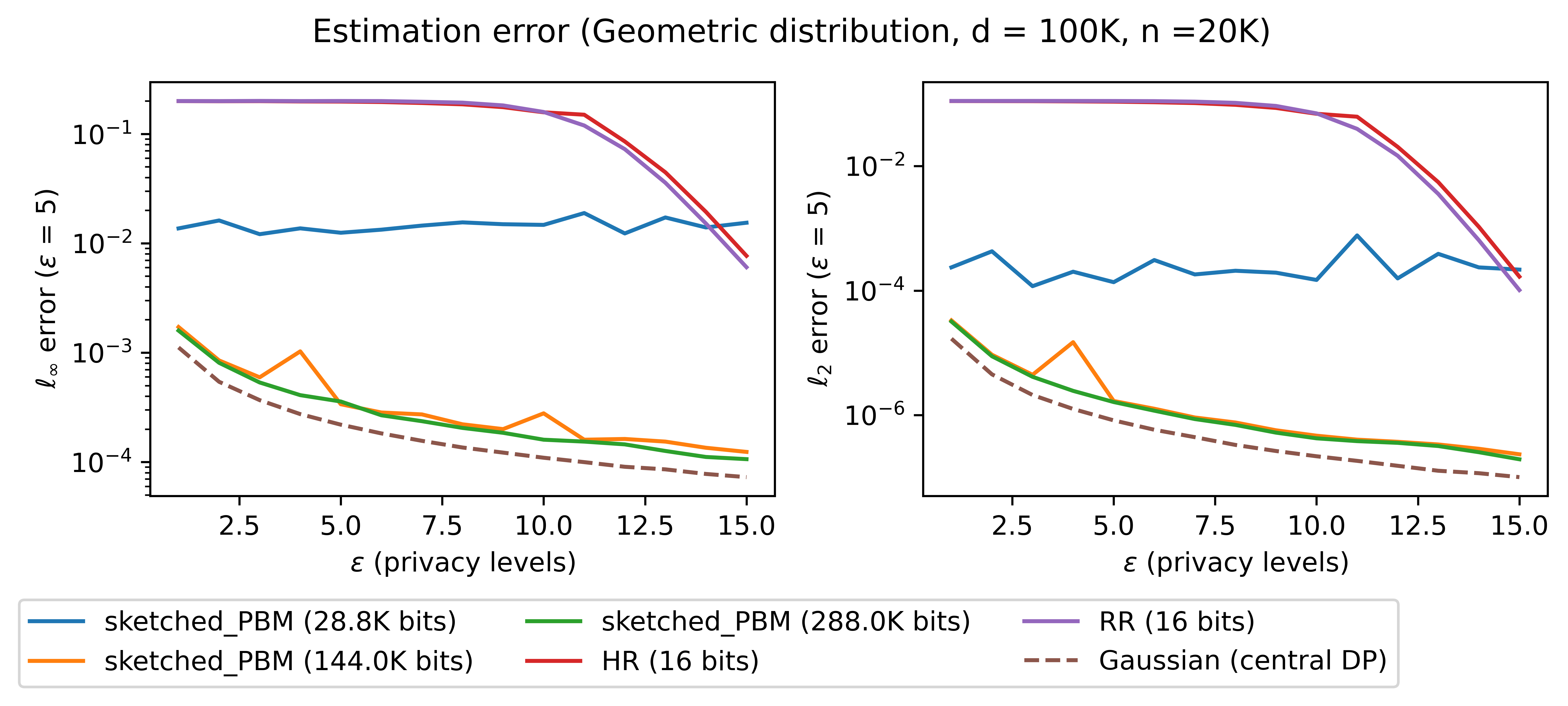}
         \includegraphics[width=0.95\textwidth]{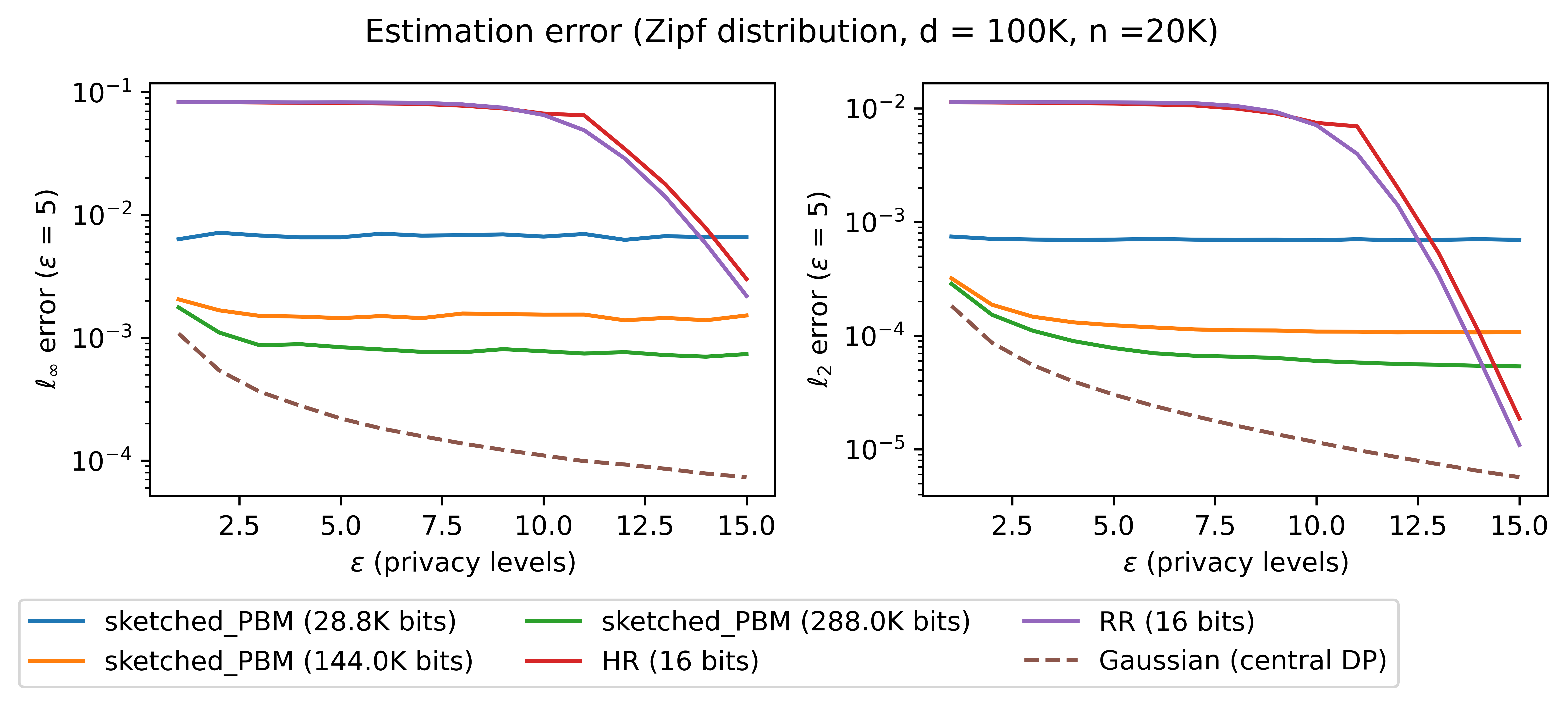}
     \caption{$\ell_\infty$ and $\ell_2$ loss with $\varepsilon = [1, 15]$.}\label{fig3}
\end{figure}

In the next set of experiments (Figure~\ref{fig3}), we fix $d = 10^5$ and $n = 2\cdot 10^4$ and vary $\varepsilon \in [1, 15]$. We compare the $\ell_2$ and $\ell_\infty$ error from different mechanisms under the Geometric distribution and Zipf distribution. We see that the sketched PBM mechanism significantly outperforms local DP mechanisms in high-privacy regime.

\newpage
\section{Sparse Private Frequency Estimation}\label{appendix:sparse_fe}

Finally, we briefly discuss the sparse frequency estimation setting, where the true histogram is assumed to be $s$-sparse $\lV \mu\lp X^n \rp \rV_0 \leq s$ for some $s \ll n \ll d$ (i.e., $X_i$ belongs to a size-$s$ subset of $[d]$). 
When $s$ is known ahead of time, the server can generate a sketch matrix $S$ according to $s$ instead of $n$, and all the analysis carries through with $n$ replaced by $s$. This improves both the communication cost and the $\ell_2$ estimation error. 

On the other hand, if $s$ is unknown but we are allowed to run a protocol with multiple rounds 
(this may or may not be possible in federated analytic settings where users may frequently drop out), 
we can first estimate $s$ (subject to privacy and security constraints) via a private $F_0$ sketch (using, for example, \citep{choi2020differentially}). 
In the second round, we can set the size of the count-sketch in Algorithm~\ref{alg1} according to $\hat{s}$. 
The communication cost of estimating $s$ is negligible compared to that of estimating $\hat{\mu}$, and hence we can still replace the dependency on $n$ with $s$ in our results.

Finally, we note that if interaction (multiple-rounds) is not allowed and  $s$ is unknown, we cannot reduce the communication from linear in $n$ to $s$. However, the thresholding trick used in the proof of Theorem~\ref{thm:private_secure_fe_achievability} can still be applied (which does not require  knowledge of $s$) and hence the $\ell_2$ error can be reduced to $O\lp \frac{s\log^2 d}{\varepsilon^2} \rp$.

\section{Omitted Proofs in the Main Body}

\subsection{Proof of Theorem~\ref{thm:l1_linear_inverse}}\label{proof:l1_linear_inverse}

Recall that $\mcal{H}_n \eqDef \lbp \mu \in \mbb{Z}_+^d \mv \lV \mu \rV_1 = n \rbp$ is the collection of all $n$-histograms. Then \eqref{eq:S_invertible} is the same as 
\begin{equation}\label{eq:S_invertible_2}
    \forall \Delta\mu \in \Delta\mcal{H}_n, \, S\cdot \Delta\mu \neq 0,
\end{equation}
where $\Delta\mcal{H}_n = \mcal{H}_n - \mcal{H}_n \eqDef \lbp \mu_1 - \mu_2 \mv \mu_1, \mu_2 \in \mcal{H}_n, \mu_1\neq \mu_2 \rbp$. Note that for any $\Delta\mu \in \Delta \mcal{H}_n$, we must have (1) $\Delta\mu_j \in \mbb{Z}^d$; (2) $\sum_j \Delta\mu_j = 0$; and (3) $\lV \Delta\mu_j \rV_1 \leq 2n$.

To show that \eqref{eq:S_invertible_2} holds when $m$ is large enough, we construct $S$ in the following probabilistic way:
$$ \forall i \in [m], j \in [d], \, S_{ij} \diid \msf{Bern}(1/2).$$
We denote the resulting probability distribution over all possible $S$ as $Q$. In addition, let $s_i \in \mbb{R}^d$ be the $i$-th row of $S$, i.e., $S = [s_1, s_2, ..., s_m]^\intercal$. Then, to prove \eqref{eq:S_invertible_2} holds for some $S$, it suffices to show 
\begin{align*}
    \Pr_Q\lbp \forall \Delta\mu \in \Delta\mcal{H}_n, \, S\cdot \Delta\mu = 0 \rbp < 1,
\end{align*}
as long as $m = O\lp n\log d /\log n \rp$, where the probability is taken with respect to the randomization over $S$.

To this end, observe that 
\begin{align}\label{eq:p_upper_bound_1}
    \Pr_Q\lbp \forall \Delta\mu \in \Delta\mcal{H}_n, \, S\cdot \Delta\mu = 0 \rbp 
    \overset{\text{(a)}}{\leq} \sum_{\Delta \mu \in \Delta\mcal{H}_n} \Pr_Q\lbp S\cdot \Delta\mu = 0 \rbp 
    \overset{\text{(b)}}{=} \sum_{\Delta \mu \in \Delta\mcal{H}_n} \lp \Pr_Q\lbp s_1 \cdot \Delta\mu = 0 \rbp\rp^m,
\end{align}
where (a) is due to the union bound, and (b) holds since each row of $S$ is generated i.i.d.

\paragraph{Additional notation.} Before we proceed to upper bound $\Pr_Q\lbp s_1 \cdot \Delta\mu = 0 \rbp$, we introduce some necessary notations. Let $\Delta\mu^+$ be the positive part of $\Delta\mu$, i.e., $\Delta\mu^+_j \eqDef \min(\Delta\mu_j, 0)$ for $j \in [d]$. Similarly,  $\Delta\mu^-_j \eqDef \min(-\Delta\mu_j, 0)$ (so we must have $\Delta\mu = \Delta\mu^+ - \Delta\mu^-$). 

For a vector $\nu \in \mbb{Z}^d$, let $\iota\lp \nu \rp$ be the multi-set containing all the \emph{non-zero} values of $\nu$. Let $\lba \iota\lp \nu \rp\rba$ be the (multi-set) cardinality of $\iota\lp \nu \rp$. 
For instance, if $\nu = [0, 1, 3, 3, 2]$, then $\iota\lp \nu \rp = \lbp 1, 2, 3, 3 \rbp$ and  $|\iota\lp \nu \rp| = |\lbp 1, 2, 3, 3 \rbp| = 4$.

Finally, let $\msf{sum}(\Delta\mu^+)$ be the set of all possible partial sums of $\iota(\Delta\mu^+)$, i.e., $\msf{sum}(\Delta\mu^+) = \lbp v\cdot \Delta\mu^+ \mv v \in \{ 0, 1\}^d \rbp$. Similarly, $\msf{sum}(\Delta\mu^-) = \lbp v\cdot \Delta\mu^- \mv v \in \{ 0, 1\}^d \rbp$.

\begin{claim}\label{claim:p_upper_bound}
For any $\Delta\mu \in \Delta\mcal{H}_n$, $\Pr_Q\lbp s_1 \cdot \Delta\mu = 0 \rbp \leq \sqrt{\frac{\pi}{2}\left\lceil\frac{\lba \iota(\Delta\mu) \rba}{2}\right\rceil}^{-1}$.
\end{claim}
\textbf{Proof of claim. } Observe that
\begin{align}\label{eq:sum_Delta_minus_plus}
    \Pr_Q\lbp s_1 \cdot \Delta\mu = 0 \rbp 
    & \overset{\text{(a)}}{=} \Pr_Q\lbp s_1 \cdot \Delta\mu^+ = s_1 \cdot \Delta\mu^- \rbp\\
    & = \sum_{\ell \in \msf{sum}(\Delta\mu^-) \cup \msf{sum}(\Delta\mu^+)} \Pr_Q\lbp s_1 \cdot \Delta\mu^+ = \ell \cap  s_1 \cdot \Delta\mu^- = \ell \rbp \nonumber\\
    & \overset{\text{(b)}}{=} \sum_{\ell \in \msf{sum}(\Delta\mu^-) \wedge \msf{sum}(\Delta\mu^+)} \Pr_Q\lbp s_1 \cdot \Delta\mu^+ = \ell \rbp \cdot \Pr_Q \lbp s_1 \cdot \Delta\mu^- = \ell \rbp \nonumber\\ 
    & \leq \max_{\ell \in \msf{sum}(\Delta\mu^+)}\Pr_Q\lbp s_1 \cdot \Delta\mu^+ = \ell \rbp,
\end{align}
where (a) holds since $\Delta \mu = \Delta\mu^+ - \Delta\mu^-$, (b) holds since $\Delta\mu^+$ and $\Delta\mu^-$ have disjoint supports and that each coordinate of $s_1$ is generated independently. 
Similarly, by symmetry, we have $\Pr_Q\lbp s_1 \cdot \Delta\mu = 0 \rbp  \leq \max_{\ell \in \msf{sum}(\Delta\mu^-)}\Pr_Q\lbp s_1 \cdot \Delta\mu^- = \ell \rbp$, so 
\begin{equation}\label{eq:upper_bound_two_mins}
    \Pr_Q\lbp s_1 \cdot \Delta\mu = 0 \rbp  \leq \min\lp\max_{\ell \in \msf{sum}(\Delta\mu^+)}\Pr_Q\lbp s_1 \cdot \Delta\mu^+ = \ell \rbp, \max_{\ell \in \msf{sum}(\Delta\mu^-)}\Pr_Q\lbp s_1 \cdot \Delta\mu^- = \ell \rbp\rp.
\end{equation}

Therefore, it remains to upper bound  $\max_{\ell \in \msf{sum}(\Delta\mu^+)}\Pr_Q\lbp s_1 \cdot \Delta\mu^+ = \ell \rbp$. To this end, observe that since each coordinate of $s_1$ is i.i.d. $\msf{Bern}(1/2)$,
$$ \Pr_Q\lbp s_1 \cdot \Delta\mu^+ = \ell \rbp = \lba \lbp v \mv v \in \{0, 1\}^d,  v \cdot \Delta\mu^+ = \ell\rbp \rba\cdot 2^{-d} = \lba \lbp A \mv  A \in 2^{\iota(\Delta\mu^+)}, \sum_{a \in A} a =\ell\rbp \rba\cdot 2^{-|\iota(\Delta\mu^+)|}, $$
where $2^{\iota(\Delta\mu^+)}$ denotes the power set of the multi-set $\iota(\Delta\mu^+)$. Notice that for the multi-set $\iota(\Delta\mu^+)$, we treat each element as a different one even some of them may possess the same value, so the cardinality of $2^{\iota(\Delta\mu^+)}$ is $2^{|\iota(\Delta\mu^+)|}$.

Now, observe that $\mcal{F}_\ell \eqDef \lbp A \mv  A \in 2^{\iota(\Delta\mu^+)}, \sum_{a \in A} a =\ell\rbp $ must form a Sperner family \citep{sperner1928satz, lubell1966short}, that is, for any $A_1, A_2 \in \mcal{F}_\ell$, neither $A_1 \subset A_2$ nor $A_2 \subset A_1$ holds. This is because otherwise, if $A_1 \subset A_2$, we must have $\sum_{A_2} a > \sum_{A_1} a$, and thus at least one of them must be not equal to $\ell$. Therefore, applying Sperner's theorem \citep{sperner1928satz, lubell1966short}, we must have 
$$  \lba \lbp A \mv  A \in 2^{\iota(\Delta\mu^+)}, \sum_{a \in A} a =\ell\rbp \rba \leq {\lba \iota(\Delta\mu^+) \rba \choose \left\lceil \frac{\lba \iota(\Delta\mu^+) \rba}{2}\right\rceil}, $$
which implies 
\begin{align*}
    \Pr_Q\lbp s_1 \cdot \Delta\mu^+ = \ell \rbp \leq {\lba \iota(\Delta\mu^+) \rba \choose \left\lceil \frac{\lba \iota(\Delta\mu^+) \rba}{2}\right\rceil}\cdot 2^{-\lba \iota(\Delta\mu^+) \rba}
    \leq \sqrt{\frac{\pi}{2}\lba \iota(\Delta\mu^+) \rba}^{-1},
\end{align*}
where the last inequality is due to basic combinatorial fact \citep[Chapter~17]{cover1999elements}. Similarly, by symmetry, we also have
\begin{align*}
    \Pr_Q\lbp s_1 \cdot \Delta\mu^- = \ell \rbp 
    \leq \sqrt{\frac{\pi}{2}\lba \iota(\Delta\mu^-) \rba}^{-1},
\end{align*}
and hence plugging in \eqref{eq:upper_bound_two_mins} we obtain
\begin{align*}
    \Pr_Q\lbp s_1 \cdot \Delta\mu = 0 \rbp  \leq \min\lp\sqrt{\frac{\pi}{2}\lba \iota(\Delta\mu^+) \rba}^{-1}, \sqrt{\frac{\pi}{2}\lba \iota(\Delta\mu^-) \rba}^{-1}\rp \leq \sqrt{\frac{\pi}{2}\left\lceil\frac{\lba \iota(\Delta\mu) \rba}{2}\right\rceil}^{-1},
\end{align*}
where the last inequality holds since $$\max\lp \lba \iota(\Delta\mu^-) \rba, \lba \iota(\Delta\mu^+) \rba \rp \geq \left\lceil\frac{\lba \iota(\Delta\mu^-) \rba+\lba \iota(\Delta\mu^+) \rba}{2}\right\rceil = \left\lceil\frac{\lba \iota(\Delta\mu) \rba}{2}\right\rceil.$$
\hfill $\square$

Now, with Claim~\ref{claim:p_upper_bound}, we proceed to bound \eqref{eq:p_upper_bound_1} as follows: 
\begin{align}\label{eq:p_upper_bound_2}
    \Pr_Q\lbp \forall \Delta\mu \in \Delta\mcal{H}_n, \, S\cdot \Delta\mu = 0 \rbp 
    & \leq \sum_{\Delta \mu \in \Delta\mcal{H}_n} \lp \Pr_Q\lbp s_1 \cdot \Delta\mu = 0 \rbp\rp^m \nonumber\\
    & \leq \sum_{\Delta \mu \in \Delta\mcal{H}_n}\sqrt{\frac{\pi}{2}\left\lceil\frac{\lba \iota(\Delta\mu) \rba}{2}\right\rceil}^{-m} \nonumber\\
    & = \sum_{\ell = 1}^{2n} \sum_{\Delta \mu : | \iota(\Delta\mu)| = \ell}\sqrt{\frac{\pi}{2}\left\lceil\frac{\ell}{2}\right\rceil}^{-m} \nonumber\\
    & \leq \sum_{\ell = 1}^{2n} {d \choose \ell}\lp 2n+1 \rp^{\ell}\sqrt{\frac{\pi}{2}\left\lceil\frac{\ell}{2}\right\rceil}^{-m} \nonumber\\
    & = \sum_{\ell = 1}^{n^*} {d \choose \ell}\lp 2n+1 \rp^{\ell}\lp{\frac{\pi}{2}\left\lceil\frac{\ell}{2}\right\rceil}\rp^{-m/2} + 
    \sum_{\ell = n^*}^{2n} {d \choose \ell}\lp 2n+1 \rp^{\ell}\lp{\frac{\pi}{2}\left\lceil\frac{\ell}{2}\right\rceil}\rp^{-m/2},
\end{align}
where $n^* \in [n]$ is a tuning parameter that will be specified later. Now, we bound the last two terms separately. For the first term, we have
\begin{align*}
    \sum_{\ell = 1}^{n^*} {d \choose \ell}\lp n+1 \rp^{\ell}\lp{\frac{\pi}{2}\left\lceil\frac{\ell}{2}\right\rceil}\rp^{-m/2}
    &\leq \lp 2n+1 \rp^{n^*}\lp{\frac{\pi}{2}}\rp^{-m/2}\sum_{\ell = 1}^{n^*} {d \choose \ell}\\
    &\leq \lp 2n+1 \rp^{n^*}\lp{\frac{\pi}{2}}\rp^{-m/2}(d+1)^{n^*+1}\\
    &\leq \exp\lp (n^*+1)\log(d+1) + n^*\log(2n+1) - \frac{m}{2}\log\lp \pi/2 \rp\rp \ra 0,
\end{align*}
as long as $m = \Omega\lp n^*\log(d+1) + n^*\log(2n+1) \rp = \Omega\lp n^*\log d \rp$ (since $n \ll d$). For the second term, observe that
\begin{align*}
    \sum_{\ell = n^*}^{2n} {d \choose \ell}\lp n+1 \rp^{\ell}\lp{\frac{\pi}{2}\left\lceil\frac{\ell}{2}\right\rceil}\rp^{-m/2}
    & \leq (2n+1)^{2n}\lp \frac{\pi n^*}{4} \rp^{-m/2} \lp \sum_{\ell=n^*}^{2n} {d \choose \ell}\rp \\
    & \leq (2n+1)^{2n}\lp \frac{\pi n^*}{4} \rp^{-m/2} \lp \sum_{\ell=0}^{2n} {d \choose \ell}\rp \\
    & = \exp\lp 2n \log(2n+1) + 2n\log(d+1) -\frac{m}{2}\lp \log n^* + \log\lp \pi/4 \rp \rp\rp.
\end{align*}
Therefore, as long as $ m = \Omega\lp \frac{2n\log(2n+1)+n\log(d+1)}{\log n^* +\log\lp \pi/4\rp} \rp = \Omega\lp \frac{2n\log d}{\log n^* +\log\lp \pi/4\rp} \rp$.

Putting both upper bounds on $m$ together, and select $n^* = \lceil n/\log n + 3 \rceil$, we conclude that as long as 
$$ m = \Omega\lp \max\lp \frac{n\log d}{\log n}+3 \log d, \frac{n \log d}{ \log n - \log\log n +3 -\log\lp \pi/4 \rp} \rp \rp = \Omega\lp n \log d /\log n \rp, $$
then $\Pr_Q\lbp \forall \Delta\mu \in \Delta\mcal{H}_n, \, S\cdot \Delta\mu = 0 \rbp \ra 0$, which implies that there must exists a feasible $S$ that distinguish all elements in $\Delta\mcal{H}_n$. 

\hfill $\blacksquare$

\subsection{Proof of Lemma~\ref{lemma:noiseless_lower_bound}}
First of all, observe that for any $\mathcal{D} \subset [n]$ such that $\mathcal{D} \leq d$
\begin{align*}
    & I\left( X_{[n]}; Y_{[n]\setminus \mathcal{D}}, h\left( \theta_{[n]}, \mathcal{D} \right) \right) \\
    &= I\left( \sum_{i\in [n]\setminus \mathcal{D}} X_i; Y_{[n]\setminus \mathcal{D}}, h\left( \theta_{[n]}, \mathcal{D} \right)\right) + I\left( Y_{[n]\setminus \mathcal{D}}, h\left( \theta_{[n]}, \mathcal{D} \right); X_{[n]} \middle\vert \sum_{i \in [n]\setminus \mathcal{D}} X_i \right) \\
    & = I\left( \sum_{i\in [n]\setminus \mathcal{D}} X_i; Y_{[n]\setminus \mathcal{D}}, h\left( \theta_{[n]}, \mathcal{D} \right)\right) \\
    & = H\left( \sum_{i\in [n]\setminus \mathcal{D}}  X_i\right) - H\left( \sum_{i\in [n]\setminus \mathcal{D}}  X_i \middle\vert Y_{[n]\setminus \mathcal{D}}, h\left( \theta_{[n]}, \mathcal{D} \right)\right)\\
    & = H\left( \sum_{i\in [n]\setminus \mathcal{D}}  X_i\right),
\end{align*}
where the second equality holds due to \eqref{S1} and the third equality holds since \eqref{C1} implies $$H\left( \sum_{i\in [n]\setminus \mathcal{D}}  X_i \middle\vert Y_{[n]\setminus \mathcal{D}}, h\left( \theta_{[n]}, \mathcal{D} \right)\right) = 0. $$ On the other hand, let $\mathcal{D'} \subset [n]$ be such that $|\mathcal{D'}| = d$ and let $j \in \mathcal{D}\setminus j$ we also have
\begin{align}\label{eq:lossless_mi_upperbound_dropout}
    &I\left( X_{[n]}; Y_{[n]\setminus \mathcal{D'}}, \theta_{\mathcal{D}'} \right) \nonumber\\ 
    &= I\left(  Y_{[n]\setminus \left\{ \{j\} \vee \mathcal{D'}\right\}}, \theta_{\mathcal{D'}\vee \{j\}} ; X_{[n]} \right) + I\left(  Y_{j} ; X_{[n]} \middle\vert Y_{[n]\setminus \left\{ \{j\} \vee \mathcal{D'}\right\}}, \theta_{\mathcal{D'}\vee \{j\}}  \right) \nonumber\\
    &= I\left(  Y_{j} ; X_{[n]} \middle\vert Y_{[n]\setminus \left\{ \{j\} \vee \mathcal{D'}\right\}}, \theta_{\mathcal{D'}\vee \{j\}}  \right) \nonumber \\
    &= H\left(  Y_{j} \middle\vert Y_{[n]\setminus \left\{ \{j\} \vee \mathcal{D'}\right\}}, \theta_{\mathcal{D'}\vee \{j\}} \right) -  H\left(  Y_{j} \middle\vert Y_{[n]\setminus \left\{ \{j\} \vee \mathcal{D'}\right\}}, \theta_{\mathcal{D'}\vee \{j\}}, X_{[n]} \right) \nonumber\\
    & \leq H\left(  Y_{j} \middle\vert Y_{[n]\setminus \left\{ \{j\} \vee \mathcal{D'}\right\}}, \theta_{\mathcal{D'}\vee \{j\}} \right) \nonumber\\
    & \leq H\left(  Y_{j} \right),
\end{align}
where the second equality is due to \eqref{S2}.
\hfill $\blacksquare$

\subsection{Proof of Lemma~\ref{lemma:lossy_lower_bound}}
Notice that by \eqref{eq:lossless_mi_upperbound_dropout}, we have $H(Y_i) \geq I\left( X_{[n]}; Y_{[n]\setminus\mathcal{D}}, h\left( \theta_{[n]}, \mathcal{D} \right) \right)$. Therefore, it suffices to lower bound $I\left( X_{[n]}; Y_{[n]\setminus\mathcal{D}}, h\left( \theta_{[n]}, \mathcal{D} \right) \right)$ subject to \eqref{C1prime}, \eqref{S1}, and \eqref{S2}. Using \eqref{S1}, we have 
$$ I\left( X_{[n]}; Y_{[n]\setminus\mathcal{D}}, h\left( \theta_{[n]}, \mathcal{D} \right) \right) = I\left( \sum_{i\in[n]\setminus\mathcal{D}} X_i; Y_{[n]\setminus\mathcal{D}}, h\left( \theta_{[n]}, \mathcal{D} \right) \right) \geq I\left( \sum_{i\in[n]\setminus\mathcal{D}} X_i; Y_{[n]\setminus\mathcal{D}}\right).$$ 

Constrained on \eqref{C1prime}, this quantity is lower bounded by $R(\beta)$.
\hfill $\blacksquare$

\subsection{Proof of Corollary~\ref{cor:exact_lower_bound}}\label{proof:exact_lower_bound}
Let $\mcal{H}_n \eqDef \lbp \lp n_1, n_2,...,n_d \rp \mv \sum_{j=1}^d n_j = n, n_j \in \mbb{Z}_+ \rbp$ be the collection of all $n$-histograms (over a size-$d$ domain). To construct a worst-case prior $\pi_{X^n}$ over $\mcal{X}^n$ such that $H\lp \sum_{i=1}^n X_i \rp = H\lp \mu\lp X^n \rp \rp$ is maximized, it suffices to find a $\pi_{\mu}$ over $\mcal{H}_n$ that has large entropy. This is because one can generate $\pi_{X^n}$ according to the following compound procedure such that $\sum_i X_i$ has marginal distribution $\pi_\mu$: first select $\mu \sim \pi_{\mu}$ and then draw $X_i$ from histogram $\mu$ without replacement.

To this end, we simply set $\pi_{\mu} = \msf{uniform}\lp \mcal{H}_n \rp$. The entropy is thus given by
\begin{align*}
H\lp \mu\lp X^n \rp \rp = \log\lba \mcal{H}_n \rba = \log\lp {d+n-1 \choose n-1} \rp = \Omega\lp n\log\lp \frac{d+n-1}{n-1} \rp \rp = \Omega\lp n\log d\rp,
\end{align*}
where the last equality holds when $d \gg n$.

\subsection{Proof of Lemma~\ref{lemma:error_function}}\label{proof:error_function}

Note that characterizing the rate function $R(\beta)$ (i.e., solving \eqref{eq:rate_function}) is equivalent to solving the following dual form:
\begin{equation}\label{eq:error_function}
\msf{err}(b) \eqDef
    \begin{pmatrix}
    & \min_{P_{Y^n |\mu\lp X^n \rp}} &\min_{\hat{\mu}} \E\lb \ell\lp \hat{\mu}\lp Y^n  \rp, \mu\lp X^n \rp \rp \rb \\
    &  \textrm{subject to } & I\lp Y^n ; \mu\lp X^n \rp\rp \leq b.
    \end{pmatrix}
\end{equation}
The dual form can be interpreted as the minimum distortion (under loss function $\ell\lp \cdot \rp$) subject to a $b$-bit communication constraint. Moreover, since $\hat{\mu}\lp \cdot \rp$ can be any arbitrary (measurable) function of $Y^n$, we suppress its dependency on $Y^n$ and simplify \eqref{eq:error_function} to
\begin{equation}\label{eq:error_function_1}
\msf{err}(b) \eqDef
    \begin{pmatrix}
    & \min_{P_{\hat{\mu} |\mu\lp X^n \rp}} &\min_{\hat{\mu}} \E\lb \ell\lp \hat{\mu}, \mu\lp X^n \rp \rp \rb \\
    &  \textrm{subject to } & I\lp \hat{\mu} ; \mu\lp X^n \rp\rp \leq b.
    \end{pmatrix}
\end{equation}

To obtain the lower bound on $\msf{err}(b)$, our strategy is to construct a hard prior distribution $\pi_{X^n}$. Following the same argument as in Corollary~\ref{cor:exact_lower_bound}, it suffices to construct a prior $\pi_{\mu}$ over $\mcal{H}_n$, such that when $\mu_1, \mu_2 \diid \pi_{\mu}$, with high-probability $\ell\lp \mu_1, \mu_2 \rp$ will be large. Once obtaining a hard $\pi_{\mu}$, we make use of the following Fano's inequality to obtain a lower bound on the smallest distortion $\E_{\mu \sim \pi_{\mu}}\lb \ell\lp \hat{\mu}, \mu \rp\rb$ one can possibly hope for.
\begin{lemma}[Fano's inequality]\label{lemma:fano}
    Let $V \sim \msf{uniform}\lp \mcal{V} \rp$ for some finite set $\mcal{V}$ and $V-U-\hat{V}$ form a Markov chain. Then
    $$ \Pr\lbp \hat{V}\lp U \rp \neq V \rbp \geq 1-\frac{I\lp U; V \rp+1}{\log\lba \mcal{V} \rba}. $$
\end{lemma}

\paragraph{Bounding the $\ell_\infty$ distortion.}
Recall that our goal is to find a prior $\pi_{\mu}$ over $\mcal{H}_n$, such that when $\mu_1, \mu_2 \diid \pi_{\mu}$, $\lV\mu_1 - \mu_2 \rV_\infty$ is large. We proceed by finding a (large) subset of $\Pi_R \subseteq \mcal{H}_n$, such that
\begin{itemize}
    \item $\lba \Pi_R \rba \geq 2^{2b}$ (where $R$ is a tuning parameter);
    \item for any $\mu_1, \mu_2 \in \Pi_R$ such that $\mu_1\neq \mu_2$, $\lV \mu_1 - \mu_2 \rV_\infty \geq \Theta\lp \frac{n\log d}{b} \rp$.
\end{itemize}
If we can find such $\Pi_R$, then by setting $\pi_{\mu} = \msf{uniform}\lp \Pi_R \rp$ and together with Fano's inequality (Lemma~\ref{lemma:fano}), we obtain
\begin{align}\label{eq:l_inf_lb_2}
\min_{\hat{\mu}} \E_{\mu}\lb \lV \hat{\mu} - \mu \rV_\infty \rb
&\geq \min_{\hat{\mu}} \E_{\mu\sim \pi_\mu}\E_{\mu}\lb \lV \hat{\mu} - \pi \rV_\infty \rb\\
& \geq \min_{\hat{\mu}} \Pr_{ \mu \sim \pi_\mu}\lbp \hat{\mu} \neq \mu \rbp \cdot \min_{\mu_1 \neq \mu_2, \mu_1, \mu_2 \in \Pi_R} \lV \mu_1 - \mu_2 \rV_\infty \\
& \geq \Theta\lp \frac{n\log d}{b} \rp \min_{\hat{\mu}} \Pr_{\mu\sim \pi_\mu}\lbp \hat{\mu} \neq \mu \rbp \\
& \overset{\text{(a)}}{\geq} \Theta\lp \frac{n\log d}{b} \rp  \lp 1- \frac{I\lp \hat{\mu};\mu \rp+1}{\log\lba \Pi_R \rba} \rp \\
& \geq \Theta\lp \frac{n\log d}{b} \rp \lp 1- \frac{b+1}{2b} \rp \\
&= \Theta\lp \frac{n\log d}{b} \rp,
\end{align}
where (a) follows from Lemma~\ref{lemma:fano}.

Therefore, it suffices to find a $\Pi_R$ that satisfies the above two criteria. To this end, consider the following construction of $\Pi_R$:
$$ \Pi_R \eqDef \lbp \lp \frac{n}{R}n_1, \frac{n}{R}n_2, ..., \frac{n}{R}n_d\rp \mv \sum_i n_i = R, n_j \in \mbb{Z}_+  \rbp.  $$
For a given $b$, we will pick $R = \Theta\lp \frac{b}{\log d} \rp$. It is then straightfoward to see that 
$$ \lba \Pi_R \rba = {d+R-1 \choose R-1} \geq \lp \frac{d+R-1}{R-1} \rp^{R-1} \geq 2^{2b}.  $$
In addition, for any distinct $\mu_1, \mu_2 \in \Pi_R$, $\lV \mu_1-\mu_2  \rV_\infty \geq \frac{n}{R} = \Theta\lp \frac{n\log d}{b} \rp$.

\paragraph{Bounding the $\ell_2$ distortion.}
We follow the same steps of analysis as in the $\ell_\infty$ case (with $\lV \cdot \rV_\infty$ being replaced by $\lV \cdot \rV^2_2$), except for requiring the set $\Pi_R$ to satisfy
\begin{itemize}
    \item $\lba \Pi_R \rba \geq 2^{\theta(b)}$;
    \item for any $\mu_1, \mu_2 \in \Pi_R$ such that $\mu_1\neq \mu_2$, $\lV \mu_1 - \mu_2 \rV^2_2 \geq \Theta\lp \frac{n^2\log d}{b} \rp$.
\end{itemize}

The construction of $\Pi_R$ under $\ell_2$ loss is slightly more involved than that in the $\ell_\infty$ case, but the central idea is to obtain a set $\Pi_R$ that matches a packing lower bound, similar to the proof of the GV bound. 

We begin with a few notations:
Let $\mcal{H}_R$ be the Hamming surface with radius $R$ (over a $d$-dimensional cube), i.e., $ \mcal{H}_R \eqDef \lbp (n_1,...,n_d)\mv \sum_{i=1}^d n_i = R, n_i \in \{0, 1\} \rbp$. Now, we construct a $\tilde{\Pi}_R \subset \mcal{H}_R$, such that for any distinct $\pi_1, \pi_2 \in \tilde{\Pi}_R$, $d_H\lp \pi_1, \pi_2 \rp \geq \frac{R}{8}$ (where $d_H(\cdot, \cdot)$ is the Hamming distance between $\pi_1$ and $\pi_2$, i.e. $\sum_{j=1}^d \bbm{1}_{\lbp \pi_1(i) \neq \pi_2(i) \rbp}$).

We claim that there exists such $\tilde{\Pi}_R$ with $\lba \tilde{\Pi}_R \rba = 2^{\theta\lp R \log d \rp}$, when $R = o(d)$. To see this, let $\tilde{\Pi}_R$ be the largest subset that satisfies the requirement. Then this would imply that for any $\pi \in \mcal{H}_R$ there exists a $\tilde{\pi} \in \tilde{\Pi}_R$, such that $d_H\lp \pi, \tilde{\pi}\rp \leq R/4$ (otherwise, one can add $\pi$ into $\tilde{\Pi}_R$ while still satisfying the requirement). This would imply the following covering bound:
\begin{equation}\label{eq:tilde_Pi_R_lower_bound}
    \lba \mcal{H}_R \rba \leq \lba \tilde{\Pi}_R \rba \cdot \lba \lbp \pi \in \mcal{H}_R: d_H\lp \pi, \tilde{\pi} \rp \leq R/4 \rbp\rba.    
\end{equation}

Now, notice that $ \lba \mcal{H}_R \rba = {d \choose R} $, and the volume of the Hamming ball can be upper bounded by
\begin{align*}
    \lba \lbp \pi \in \mcal{H}_R: d_H\lp \pi, \tilde{\pi} \rp \leq R/4 \rbp\rba
    &= \sum_{i=1}^{R/8} {d-R \choose i}{R\choose i} \\
    &\leq {d-R \choose R/8} \sum_{i=0}^{R/8} {R \choose i}\\
    & \leq d^{R/8}\cdot 2^{Rh_{\msf{b}}(1/8)},
\end{align*}
where in the last inequality we use upper bound on binomial partial sum: $\sum_{i=1}^k {R\choose k} \leq 2^{Rh_{\msf{b}}\lp \frac{k}{R} \rp}$ where $h_{\msf{b}}(\cdot)$ is the binary entropy function.

Plugging the upper bound into \eqref{eq:tilde_Pi_R_lower_bound}, we obtain
$$ \lba \tilde{\Pi}_R\rba \geq \frac{{d \choose R}}{d^{R/8}\cdot 2^{Rh_{\msf{b}}(1/8)}} = 2^{\lp R\log\lp \frac{d}{R} \rp - R\lp \frac{1}{8}+h_{\msf{b}}\lp \frac{1}{8} \rp \rp\rp} = 2^{\Theta\lp R \log d \rp} = 2^{\Theta\lp b \rp},  $$
when $d \gg R$ and $R = \Theta\lp \frac{b}{\log d} \rp$.

Finally, we rescale $\tilde{\Pi}_R$ to obtain $\Pi_R$: $\Pi_R\eqDef \lbp \frac{n}{R}\pi: \pi \in \tilde{\Pi}_R \rbp$. Obviously, we have $\lba \Pi_R\rba  = \lba \tilde{\Pi}_R\rba \geq 2^{\Theta\lp b \rp}$.
Moreover, for any distinct $\mu_1, \mu_2 \in \Pi_R$, $\lV \mu_1 - \mu_2\rV^2_2 \geq d_H\lp {\mu_1, \mu_2}\rp\cdot \frac{n^2}{R^2} = \Theta\lp \frac{n^2\log d}{b} \rp$. 
This completes the lower bound on $\msf{err}(b)$ under the $\ell_2$ loss.

\end{document}